\documentclass{elsart}
\usepackage{natbib}

\def \apj{ApJ}
\def \apjl{ApJ}
\def \aj{AJ}
\def \apjs{ApJS}
\def \aap{A\&A}
\def \mnras{MNRAS}
\def \araa{ARA\&A}
\def \apss{AP\&SS}
\def \nat{Nature}
\def \physrep{Physics Reports}

\usepackage{graphicx}

\usepackage{amssymb}

\begin{document}

\begin{frontmatter}



\title{Prospects for AGN studies with ALMA}


\author{R. Maiolino}

\address{INAF - Astronomical Observatory of Rome}

\begin{abstract}
These lecture notes provide an introduction to
mm/submm extragalactic astronomy, focused on AGN studies,
with the final goal of preparing students to 
their future exploitation of the ALMA capabilities.
I first provide an overview of the current results obtained
through mm/submm observations of galaxies and AGNs, both local and
at high redshift.
Then I summarize the main mm/submm facilities that are currently available.
ALMA is then presented with a general description and by providing some
details on its observing capabilities. Finally, I discuss some of the
scientific goals that will be achievable with ALMA in extragalactic
astronomy, and for AGN studies in particular.
\end{abstract}


\end{frontmatter}

\section{Introduction}
\label{sec_intro}

The Atacama Large Millimeter Array (ALMA) is one of the largest ground-based
astronomy projects of the next decade, which will revolutionize 
several fields of astronomy. A large community of scientists is
expected to use ALMA to tackle several outstanding questions in astrophysics.
However, mm/submm astronomy
is often considered a field restricted to experts. In the case
of students and young scientists in particular, the limited
familiarity with mm/submm facilities and observations may prevent them
to fully exploit the ALMA capabilities in the future.
These lecture notes are aimed at providing
students and young researches some background on mm/submm extragalactic astronomy,
with a focus on the investigation of AGNs. I will first provide a quick
overview of the current results obtained through extragalactic mm/submm observations,
by focusing on AGNs (\S\ref{sec_mm_astronomy}). I will then summarize the currently available (and forthcoming)
mm-submm facilities (\S\ref{sec_current_facilities}).
Then I will shortly describe ALMA and summarize its observing
capabilities (\S\ref{sec_alma}).
Finally, I will discuss some of the ALMA prospects
for extragalactic studies, and in particular for AGNs, both in the local universe and at cosmological
distances (\S\ref{sec_alma_prospects}). These lecture notes are far from being exhaustive; several scientific
cases will not be discussed at all; the main goal of these notes is only to provide an introduction to
mm/submm extragalactic astronomy and to highlight some scientific cases that ALMA will be able to tackle.

\section{Millimetric and submillimetric extragalactic astronomy}
\label{sec_mm_astronomy}

\begin{figure}
\centerline{
\includegraphics[scale=0.6,bb=44 225 580 615]{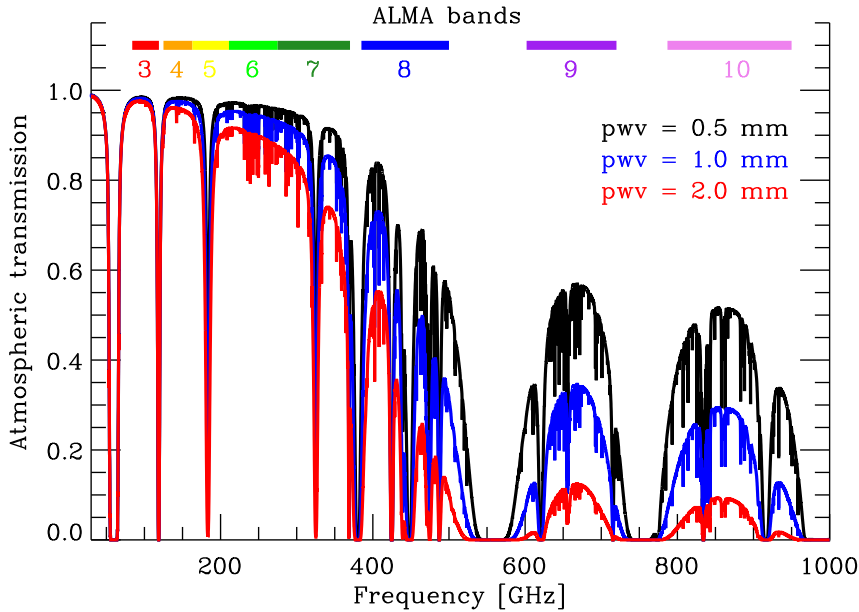}}
\caption{Atmospheric transmission at Chajnantor Plateau, the ALMA site, with different amounts
of precipitable water vapor. The horizontal colored bars
indicate the frequency ranges of the ALMA bands.
}
\label{fig_atm_transmission}
\end{figure}

This branch of astronomy includes observations at wavelengths between $\sim$10~mm and
$\sim$300~$\mu$m. Longer wavelengths are traditionally identified as radio-astronomy
domain. Shorter wavelengths, out to mid-IR wavelengths, are unobservable from ground
because of the nearly complete atmospheric absorption (although some sites, under exceptional
conditions, allow observations out to $\sim 200 \mu$m.). Even within the mm-submm range
not all wavelengths are equally easy to observe, since the sky transparency on
average decreases rapidly at shorter wavelengths. At $\rm \lambda < 700 \mu m$ only
a few atmospheric windows are accessible, and only under optimal weather conditions.
This issue is clearly illustrated in Fig.~\ref{fig_atm_transmission}, which
shows the atmospheric transmission at the ALMA site.

The main source of opacity at these wavelengths is the water vapor. This is the reason
for locating mm-submm observatories
at dry and high altitude sites, where the amount of water vapor is much reduced.
However, even at these optimal sites there are strong variations of the
the water vapor, which make the atmospheric transmission change strongly
(Fig.~\ref{fig_atm_transmission}) both
on long (seasonal) and short (day/night) time scales.

Given the difficulties of observing at these wavelengths one may
wonder why international agencies are investing so much effort to
develop facilities with enhanced observing capabilities in these bands.
The mm-submm band contains a wealth of information that cannot be inferred
from any other band. Most of the $\sim$150 molecules known so far in
the {\it cold} interstellar medium (see http://astrochemistry.net for an
updated list)
emit their rotational transitions in the mm-submm bands, with a
density of about 70~lines/GHz. All of these transitions
are important diagnostics of the chemistry, of the physics and of the dynamics
of the Inter Stellar Medium (ISM) from which stars form. Some of these lines are so strong (e.g. the
CO transitions) to be powerful tools to trace the dynamics and the gas physics even in distant galaxies.
Furthermore, some of the strongest lines emitted by the ISM of any galaxy,
such as the [CII]158$\mu$m and the [OI]63$\mu$m fine structure lines (the two main coolants of the ISM),
are redshifted into the mm-submm bands at z$>$2--4.

Within the context of the continuum emission, the mm-submm bands encompass
the Rayleigh-Jeans region of the warm dust thermal
emission (which traces star formation and the dust mass),
the high frequency tail of the synchrotron emission
(dominating the radio emission in most galaxies) and of the free-free emission
(tracing HII regions). At high redshift the prominent IR dust thermal bump (which
dominates the Spectral Energy Distribution --SED-- in starburst galaxies)
is shifted into the submm band,
therefore making this one of the best spectral regions to search and
characterize high-z star forming galaxies.

This was just a very quick glance at the scientific motivations behind the development of
mm-submm facilities, and mostly limited to the extragalactic field. Young stellar objects,
protostars and proto-planetary systems are, for instance, additional fields where the mm-submm
range is crucial for a thorough investigation.

The importance of the mm-submm band within the extragalactic context will become more
obvious in the following sections, where 
I will provide some (shallow) background on what we
currently know of external galaxies based on mm-submm observations,
and
where some extragalactic ALMA science cases will be discussed.

On the technical side, it is important to mention that the (sub)mm is currently the
shortest wavelength where sensitive, many-elements coherent detection interferometers
are feasible from the ground. These can simultaneously provide high angular resolution,
sensitivity, and image reconstruction fidelity. Direct detection interferometers at
shorter wavelengths (e.g. mid/near-IR) can achieve similar angular resolution, but are more
severely constrained in terms of sensitivity and image fidelity.

\begin{figure}
\centerline{
\includegraphics[scale=0.8]{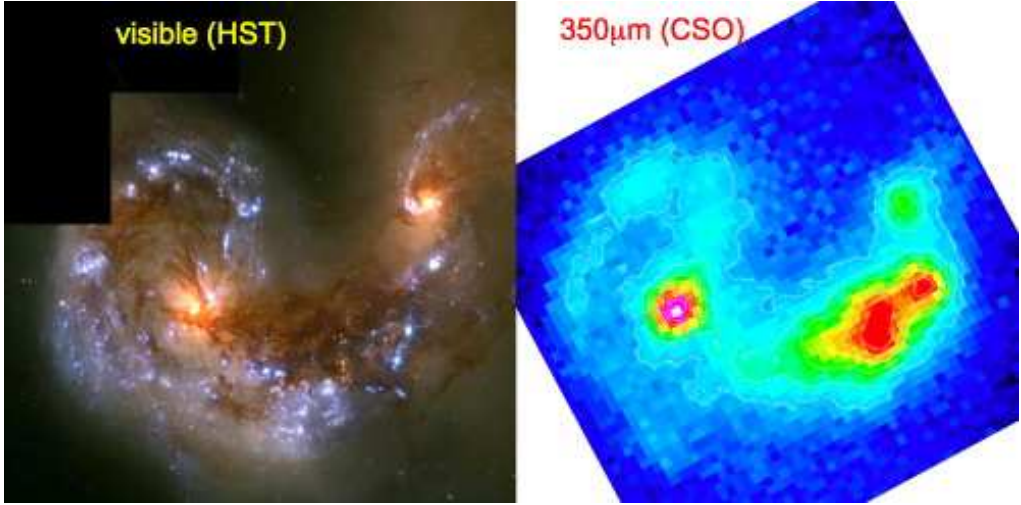}}
\caption{Visible image of the interacting galaxies ``Antennae'' ({\it left})
compared with their
submm image at 350$\mu$m ({\it right}). Note that most of the far-IR emission
comes from a region that is heavily obscured at optical wavelengths.
Credit of the Space Telescope Science Institute (optical HST image) and
of C. Dowell (submm CSO image).}
\label{fig_antennae}
\end{figure}

\subsection{Local normal and starburst galaxies}
\label{sec_past_normal_local}


The warm dust emitting at far--IR wavelengths is mostly heated by the UV radiation field
of young massive stars in star forming regions. As a consequence, the far infrared luminosity
$\rm L_{FIR}$ and its submm Rayleigh-Jeans part are considered good tracers of star formation
in galaxies. In particular, these bands are useful to trace obscured star formation,
since they are virtually unaffected by dust extinction. This is evident in Fig.~\ref{fig_antennae},
where the 350$\mu$m map of the interacting galaxies ``Antennae'' (obtained at
the CSO telescope, C. Dowell, priv. comm.)
is compared with the optical HST image: the region of most vigorous star formation traced by
the submm emission is actually the most obscured and less visible at optical wavelengths.
The main problems of the current instrumentation (bolometer arrays on single dish telescopes)
in tracing star formation in external galaxies are their limited
sensitivity and their poor angular resolution (10$''$--20$''$). Both these issues will no longer
be a problem with ALMA, which will have sensitivities orders of magnitude better and 
an angular resolution similar to HST.


As already mentioned, also most of the {\it gas} phase of the cold ISM emits in the
mm-submm range. More specifically, it is in this band that
most of the molecular gas transitions are observed. However, {\it cold} molecular hydrogen H$_2$
(by far the most abundant molecule in the cold gas phase) cannot be detected directly,
since it has no electric dipole moment (therefore rotational transitions with
$\rm \Delta J\pm 1$ are not allowed).
Carbon monoxide CO is the second most abundant molecule:
its rotational levels are excited by collisions with H$_2$, producing the brightest molecular
lines in the spectrum of any galaxy. The luminosity of the CO rotational
transitions, and in particular
the fundamental one J(1$\rightarrow$0), are widely used as tracers of the molecular gas
mass through a linear relation: $\rm L_{CO}=\alpha M(H_2)$. However, the conversion factor
$\alpha$ is found to depend on the gas metallicity as well as on its physical conditions
(temperature and density).

The critical density of the CO transitions is relatively low, at least for the
low rotational levels. For instance, the critical density of CO (1--0) (at 115 GHz)
is about $\rm 2~10^3~cm^{-3}$.
As a consequence, CO is generally little effective in tracing the high density regions of molecular
clouds. Moreover, the large optical thickness the CO lines (at least for the low rational
levels) prevents us to penetrate dense molecular regions.
Dense regions are better traced by other species, such as
HCN, HCO$^+$ and CS, which are characterized by higher
critical densities for transitions in the same frequency bands as the CO ones.
For instance HCN (1--0), HCO$^+$ (1--0) and CS (2--1) (observable in the 3mm band), have
critical densities of $\rm 2~10^6$, $\rm 1.5~10^5$ and
$\rm 4~10^5~cm^{-3}$, respectively. However, these lines
are typically one order of magnitude fainter than CO.

\begin{figure}
\centerline{
\includegraphics[scale=0.57]{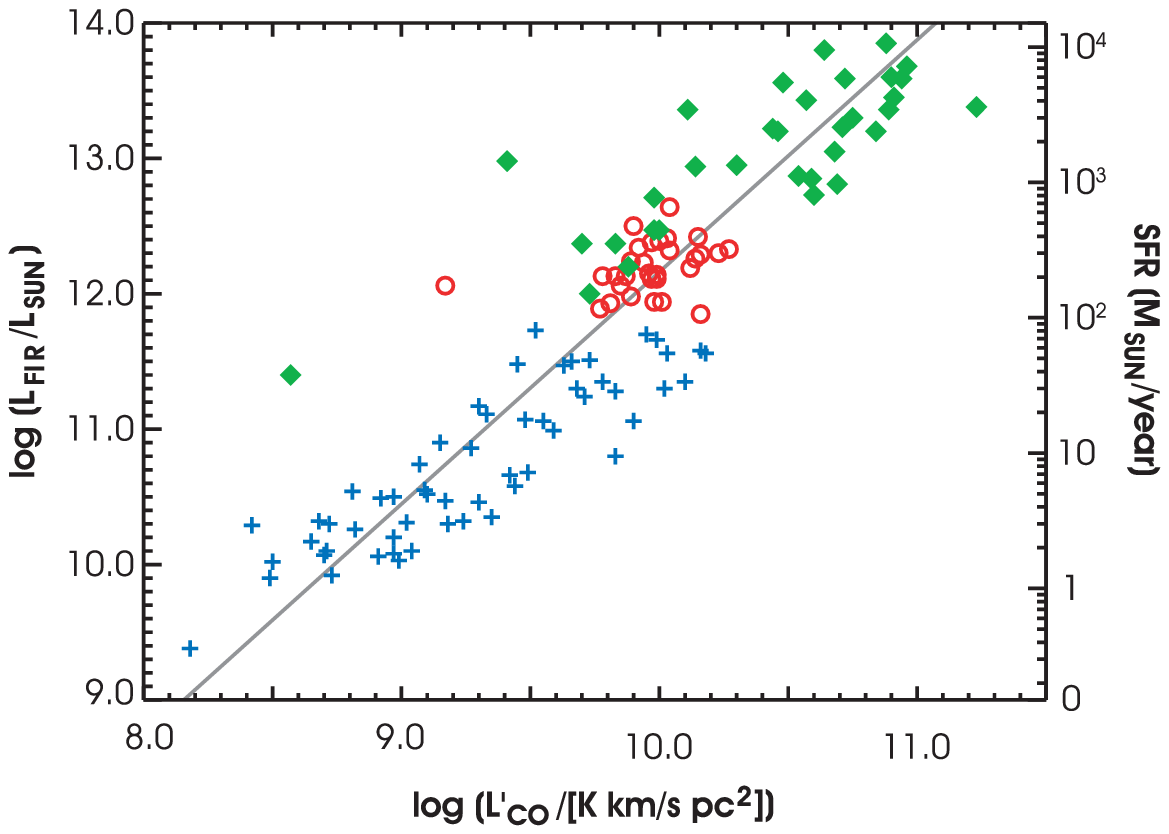}
\includegraphics[scale=0.57]{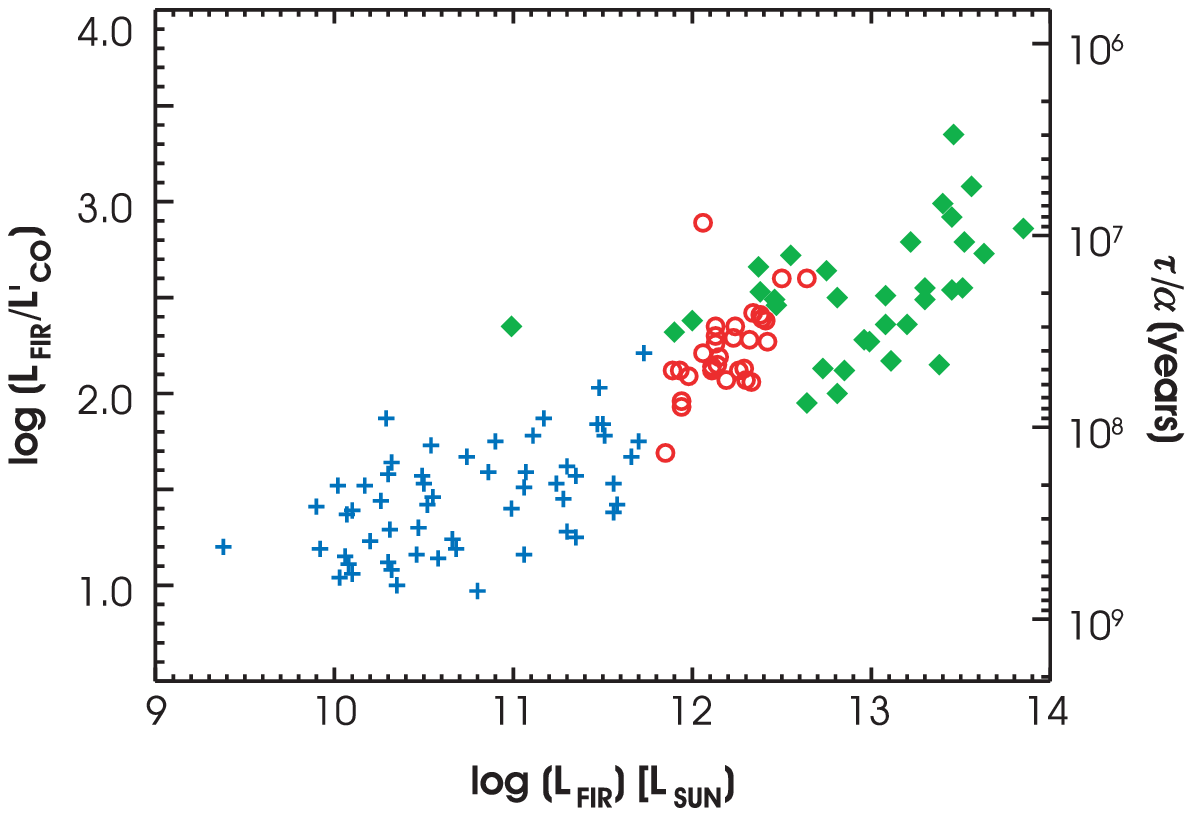}
}
\caption{{\it Left:} Relationship between far-IR luminosity and CO luminosity obtained for normal
spirals (blue crosses), ULIRGs (red circles) and high redshift galaxies with CO detections
(green diamonds). Note the non-linear relationship
$\rm \log{L_{FIR}}=1.7\log{L'_{CO}}-5.0$, indicated by the solid line. The right-hand axis
translates the far-IR luminosity into star formation rate. {\it Right:} $\rm L_{FIR}/L'_{CO}$
as a function of the far--IR luminosity. The right-hand axis gives the gas exhaustion timescale,
where $\alpha$ is the $\rm L'_{CO}$--to--H$_2$ conversion factor ($\rm \alpha \approx 1-5 ~M_{\odot}
~(K~km~s^{-1}~pc^{-2})^{-1}$). Both figures are from \cite{solomon05}.}
\label{fig_fir_co}
\end{figure}

There is a relationship between infrared luminosity and CO luminosity, which is by itself
not surprising (as most Luminosity-Luminosity relations).
More interesting is the fact that the relation is not linear, being
$\rm L_{FIR} \propto L_{CO}^{1.7}$ (Fig.~\ref{fig_fir_co}, from \cite{solomon05}).
Since $\rm L_{FIR}$ is proportional to the star formation rate SFR,
while $\rm L_{CO}$ is proportional
to the molecular gas content M(H$_2$) (which is the fuel for star formation),
the non linear relation
implies that the star formation efficiency, defined as
$\rm SFE = SFR/M(H_2)$ increases with the luminosity of the system. The inverse of the star
formation efficiency is the gas exhaustion timescale, $\rm \tau = 1/SFE\propto L_{CO}/L_{FIR}$,
i.e. the time
required by the starburst to totally consume the available molecular gas, if the star formation
proceeds at the rate currently observed. The non--linear relation between $\rm L_{FIR}$
and $\rm L_{CO}$ implies that the gas exhaustion timescale decreases with luminosity, reaching
values as low as $\rm 10^7$, as illustrated in Fig.~\ref{fig_fir_co}.
It is clear that the most powerful starbursts are also short lived. However, it should be mentioned
that this picture may be biased, especially for what concerns high redshift sources, due to
the incomplete census of galaxy populations. Indeed,
\cite{daddi08} have recently identified some {\it powerful} starburst
galaxies characterized by {\it low} SFE at z$\sim$2.

The far-IR luminosity appears to be {\it linearly} correlated with the luminosity of
dense gas tracers, such as HCN. The relation is linear over a very broad luminosity
range spanning from Galactic star forming regions to powerful starburst galaxies
\citep{wu05}. This result suggests that star formation always
occurs in dense molecular clouds, and that star formation in starburst systems may simply
be a scaled up version of Galactic star formation. However, also the linearity between
$\rm L_{FIR}$ and $\rm L_{HCN}$ has been recently questioned through indications that
the ratio between these two quantities increases with the luminosity \citep{gracia08}.
Moreover, the presence of an AGN may also affect the intensity of the HCN emission, as discussed
in \S\ref{sec_past_agn_local}.


Interferometers allowed astronomers to map the molecular gas in large samples
of galaxies, generally by exploiting the CO lines, but more recently also through transitions
of various other species. The CO emission in galaxies show a variety of morphologies:
spiral patterns, bars, rings, nuclear concentrations and irregular distributions
\citep{helfer03}.
The CO maps also reveal important kinematic information. In the case of disk or ring-like
morphologies the gas kinematics is generally dominated by (nearly) regular rotation.
In the case of barred galaxies, the molecular gas kinematics is often characterized by
prominent streaming motions along the stellar bar \citep[e.g.][]{regan99,schinnerer07}.
This is regarded
as direct evidence that the non-axisymmetric potential of a bar is effective in funneling
gas into the central region (which may eventually produce a central starburst, or be further
driven into the nuclear region to fuel an AGN).

\subsection{Local AGNs}
\label{sec_past_agn_local}


AGNs generally heat their circumnuclear dust to temperatures much higher than starburst
galaxies. Indeed, active nuclei are generally characterized by strong mid-IR and near-IR
emission, indicating dust temperatures of several hundred degrees and reaching the dust
sublimation temperature (1500--2000~K). At far-IR and submillimeter wavelengths the relative
contribution of AGN and star formation to the dust heating is debated. Spectral decomposition techniques,
as well as the correlation between far-IR emission and other tracers of star formation, suggest
that the far-IR and submm emission is dominated by star formation in the host galaxies, even
in powerful QSOs \citep{schweitzer06,netzer07}.
Disentangling {\it spatially} the far-IR/submm emission due to
the AGN from the host galaxy emission is very difficult, if not impossible, with current facilities
due to the lack of angular resolution at these wavelengths.
Disentangling the two components is
important not only to determine the contribution of AGNs to the far-IR/submm radiation, but
also to constrain models of the obscuring dusty torus, as we shall see
in \S\ref{sec_alma_local}.


Identifying the main mechanism responsible for fuelling AGNs has been one of the hottest topics
in the last decade. Non-axisymmetric potentials introduced by stellar bars and galaxy interactions
were considered as promising mechanisms, but various studies failed in finding
any excess of these morphologies in AGNs.
CO observations offer the possibility of directly witnessing the gas fuelling towards the nucleus.
Intensive campaigns have been performed with millimetric interferometers to investigate this
issue by mapping the CO emission in galaxies with different types of nuclear
activity \citep[e.g. the NUGA project, ][]{garcia05,garcia07,combes04}. The main result is that there is
no evidence for systematic differences, in terms of molecular gas distribution and kinematics,
between galaxies hosting AGNs and quiescent ones. In particular, Seyfert galaxies appear
characterized by a wide variety of molecular gas distributions: streaming motions along stellar
bars, rings, nuclear concentrations, nuclear voids and irregular distributions.
The lack of any relationship between the presence of an AGN and
the CO morphology/dynamics, indicates that there is no ubiquitous evidence for current
fuelling of the AGN. One possibility, is that the large scale fuelling phase and the AGN phase may
not be simultaneous.
Another consideration is that local AGNs have low luminosities and do not require large fuelling
rates from the host galaxies. More specifically, most local Seyfert nuclei are characterized by
black hole accretion rates of about $\rm 10^{-3}~M_{\odot}~yr^{-1}$; at these rates even a single molecular
cloud of $\rm 10^6~M_{\odot}$ can keep the nucleus active for about 1~Gyr.
The fuelling problem is more serious for powerful QSOs, where the accretion rates may exceed $\rm
1~M_{\odot}~yr^{-1}$. In QSOs
some mechanism capable of funneling molecular gas from the host galaxy into
the nuclear region is actually required. However, QSOs are much more distant than Seyfert galaxies.
The limited sensitivity and angular resolution of current mm interferometers hampers our capability of
mapping the molecular gas distribution in QSO hosts, which will instead be easy with ALMA.


\begin{figure}
\centerline{
\includegraphics[scale=0.42]{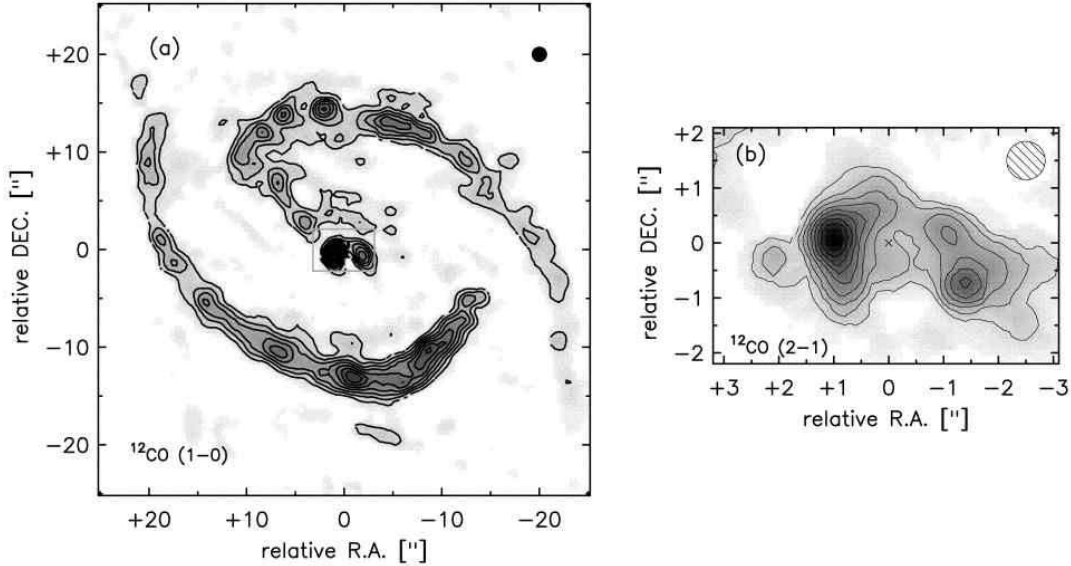}
}
\caption{{\it Left:} CO(1--0) interferometric maps of the Seyfert 2
galaxy NGC1068.
{\it Right:} Zoom of the central region of the CO(2--1) map.
The cross indicates the location of the radio nucleus.
Circles at the top-left corner indicate the beam size. At the
distance of NGC1068 one arcsec corresponds to about 80 pc.
Both maps are from \cite{schinnerer00}.}
\label{fig_n1068}
\end{figure}

NGC~1068 is one of the best studied AGNs also at mm-submm wavelengths. At the highest resolution
currently achievable the nuclear CO distribution shows a broken ring-like structure, about 2$''$ in size
(corresponding to 140~pc), as illustrated in Fig.~\ref{fig_n1068} \citep{schinnerer00}.
At the specific location of the AGN (cross in Fig.~\ref{fig_n1068}) little CO emission is observed.
Does this mean that little or no
molecular gas is present in the vicinity of the AGN? This is somewhat contrary to expectations, given
the need of gas in the vicinity of the AGN both to feed it and to provide the heavy
obscuration characterizing NGC~1068 \citep{matt97}. Actually, we know that in the vicinity
of the AGN, on scales of 1--10~pc, dense molecular gas and dust are present, based on radio
and mid-IR interferometric observations \citep{greenhill96,jaffe04}. Possibly most of the molecular
gas in the nuclear region is dense and warm, and therefore not properly sampled by the CO lines
(at least not by the low rotational transitions).
Maps of other lines tracing dense gas, such as HCN, have lower angular resolution \citep{tacconi94},
but do show a much higher concentration on the nucleus with respect to CO.
\cite{usero04} detected several molecular species in the nucleus of NGC~1068
tracing not only dense gas, but also a very complex chemistry. These findings
indicate that the AGN has created a giant ``X-ray Dominated Region'' (XDR).
Hard X-ray photons emitted by the AGN can penetrate deep into the circumnuclear molecular clouds
and keep the temperature high over an extended region.
The high gas temperature in XDR's
favors the formation of various molecular species such as HCN \citep{maloney96,lepp96,
maijerink07}. In contrast,
Photo Dissociation Regions (PDR),
which are generated by the UV photons of star forming regions, are characterized by a much
narrower region with enhanced temperature, making the production of various molecular species much less
efficient than in XDR. These fundamental differences between XDR's and PDR's suggest that
XDR-enhanced species can be used to unveil the presence of heavily obscured AGNs that escaped detection
at other wavelengths. Within this context \cite{gracia06} and \cite{kohno08}
developed a diagnostic diagram involving the line
ratios HCN/HCO$^+$ versus HCN/CO, where pure Seyfert nuclei are clearly separated from starburst
nuclei,
in the sense  that the former show enhanced HCN emission \citep[but see also caveats discussed in ][]{pap07}.
This and other complementary diagnostic diagrams will be usable with ALMA to identify obscured AGNs even in
distant galaxies.

\subsection{Distant starburst galaxies}
\label{sec_past_starburst_highz}


\begin{figure}
\centerline{
\includegraphics[scale=0.6,bb=45 205 575 670]{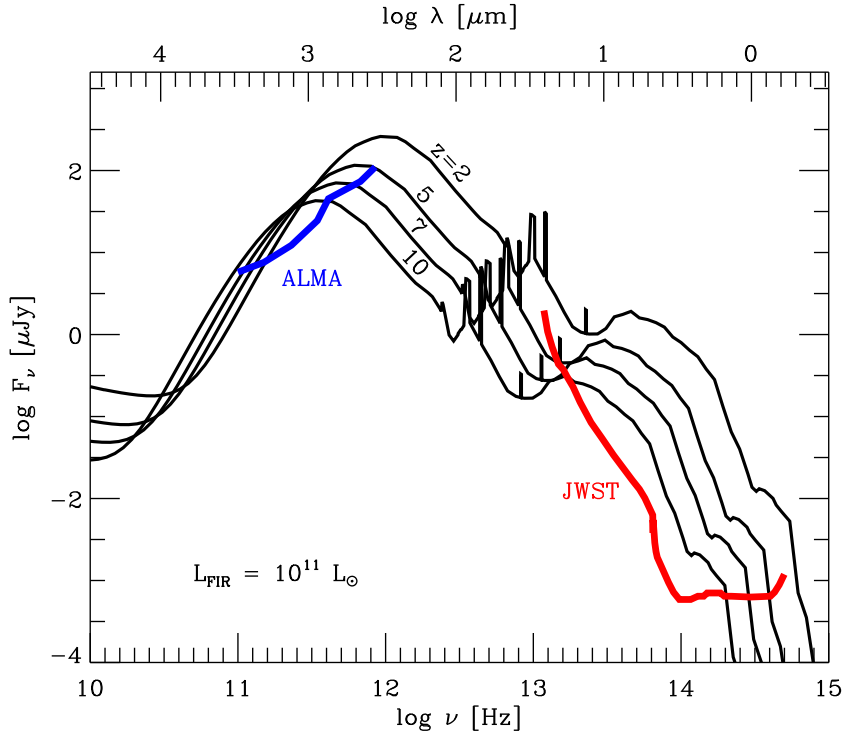}
}
\caption{Observed 
SED of a star forming galaxy with $\rm L_{FIR}=10^{11}~L_{\odot}$ (scaled from the SED of M82)
at different redshifts (z$=$2, 5, 7 and 10, as labelled). Note that the observed flux at $\rm \lambda _{obs}\sim 1~mm$ changes very little.
The thick blue and red lines show the ALMA and JWST sensitivities for continuum detection at 5$\sigma$
with an integration of $\rm 10^5~sec$.}
\label{fig_sed_z}
\end{figure}

\begin{figure}
\centerline{
\includegraphics[scale=0.6]{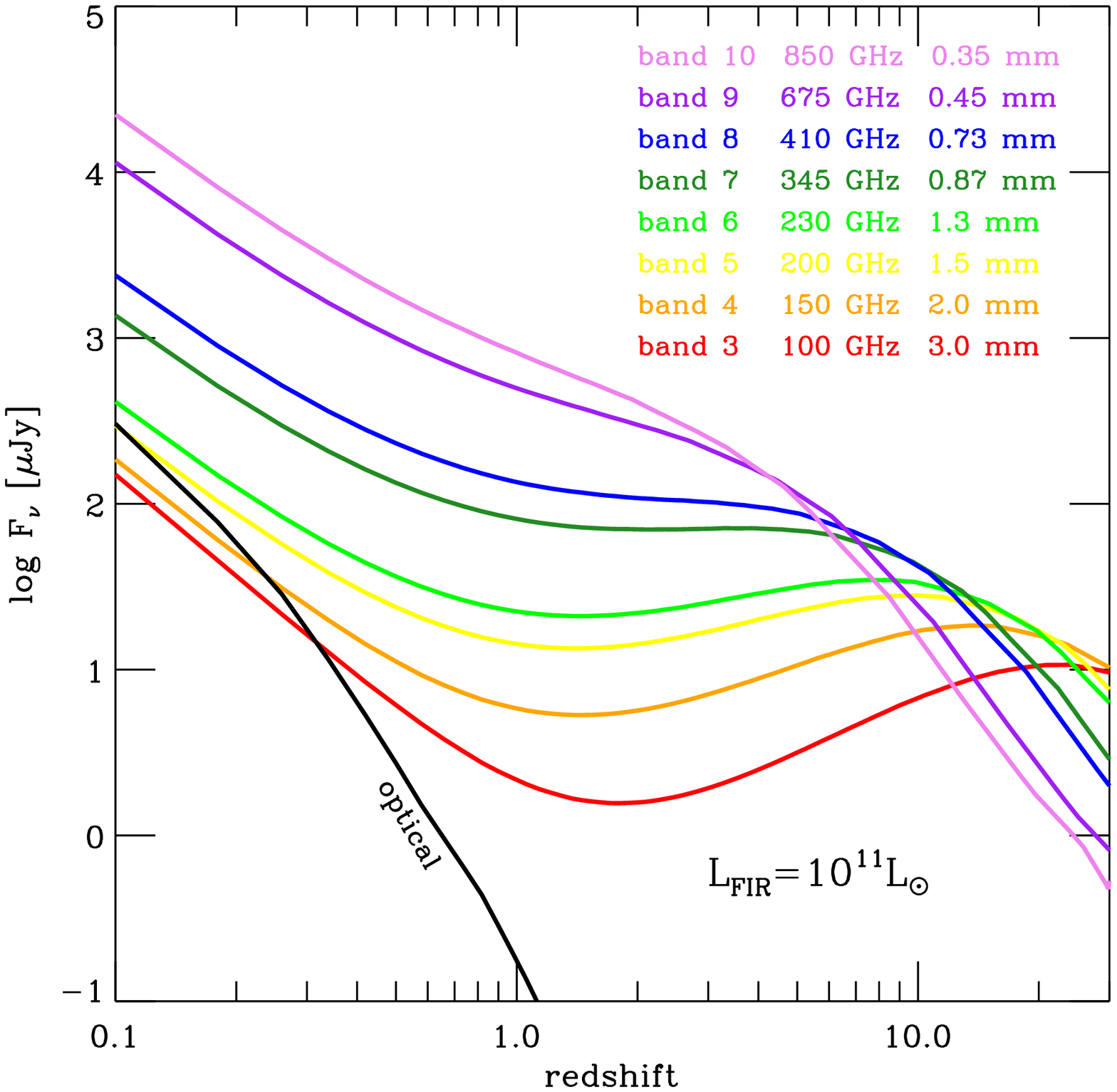}
}
\caption{Predicted flux densities of a star forming galaxy with
$\rm L_{FIR}=10^{11}~L_{\odot}$ (by scaling the M82 template) observed in the
different mm-submm ALMA bands.
Note the nearly flat trend at 1$<$z$<$10 at wavelengths close to $\sim$1~mm.
The predicted flux density in the optical ($\lambda _{obs}=5500$\AA) is also shown for
comparison.
}
\label{fig_kneg}
\end{figure}

As the spectrum of a starburst galaxy moves in redshift, the global flux is reduced according to the
1/D$^2$ cosmological dimming, but the intrinsic luminosity
observed at a fixed mm-submm wavelength increases
as a consequence of the steeply rising submm continuum.
At z$>$1 such strong negative K--correction counteracts completely the cosmological dimming, so
that detecting a source at z$=$10 is as easy as detecting a source at z$=$1 (for a given intrinsic
luminosity).
This effect is illustrated in Fig.~\ref{fig_sed_z}, which shows the observed flux
distribution of a star forming galaxy at different redshifts.
The flux observed at $\lambda  _{obs}= 1$~mm is essentially unchanged within the redshift range
$\rm z\sim 1-10$.
Fig.~\ref{fig_kneg} shows the same effect by plotting, for a given starburst template,
the observed flux at various wavelengths (and specifically in the ALMA bands)
as a function of redshift. The observed flux remains
nearly constant in the redshift range 1$<$z$<$10 at wavelengths longer than about 800$\mu$m.
At wavelengths shorter than $\sim$500$\mu$m the K--correction is not strong enough to compensate
for the cosmological dimming. At wavelengths longer than $\sim$2~mm the K-correction is strong,
but the observed flux is more than one order of magnitude fainter than observed at 1~mm; moreover,
at $\rm \lambda > 2~mm$ there is an increasing ``risk'' of contamination by non-thermal sources.

An interesting consequence of the strongly negative K--correction at mm-submm wavelengths is that
in any deep field one expects to see many more galaxies at z$>$1.5 than at lower redshifts. This
is just the opposite of what observed in the optical, where the distribution of observed fluxes is
dominated by the cosmological dimming (Fig.~\ref{fig_kneg}),
which makes any optical deep field dominated by galaxies
at z$<$1.5, and only a small fraction of galaxies at higher redshifts.


The negative K--correction allowed the discovery of a large number (a few 100) of starburst
galaxies at high-z, thanks to extensive surveys exploiting array of bolometers available
on single dish telescopes \citep[see e.g. ][for a review]{blain02,smail06}.
Dusty starburst galaxies at high-z discovered through the detection of their submm continuum
are often dubbed as Sub-Millimeter Galaxies (SMGs). Although, high-z dusty starbursts have been
also discovered through observations in the mm band, for sake of simplicity we will refer to the
whole population as ``SMGs''.

Much effort has been invested for several years in the identification of the optical (or near-IR)
counterparts of SMGs, mostly with the goal of determining their redshift through spectroscopic
followup. However, as we will discuss in \S\ref{sec_current_facilities}, the angular resolution
of single dish telescopes is so low (11$''$--18$''$) that several optical/near-IR candidate counterparts
are found within the telescope beam.
The optimal way to identify the true counterpart would be to obtain mm-submm observations at
higher angular resolution with a mm-submm interferometer. However, as discussed in
\S\ref{sec_current_facilities}, the sensitivity to continuum of current mm-submm interferometers
is low, making this approach very expensive in terms of observing time, especially if a
statistically meaningful sample is needed.

\begin{figure}
\centerline{
\includegraphics[scale=0.37]{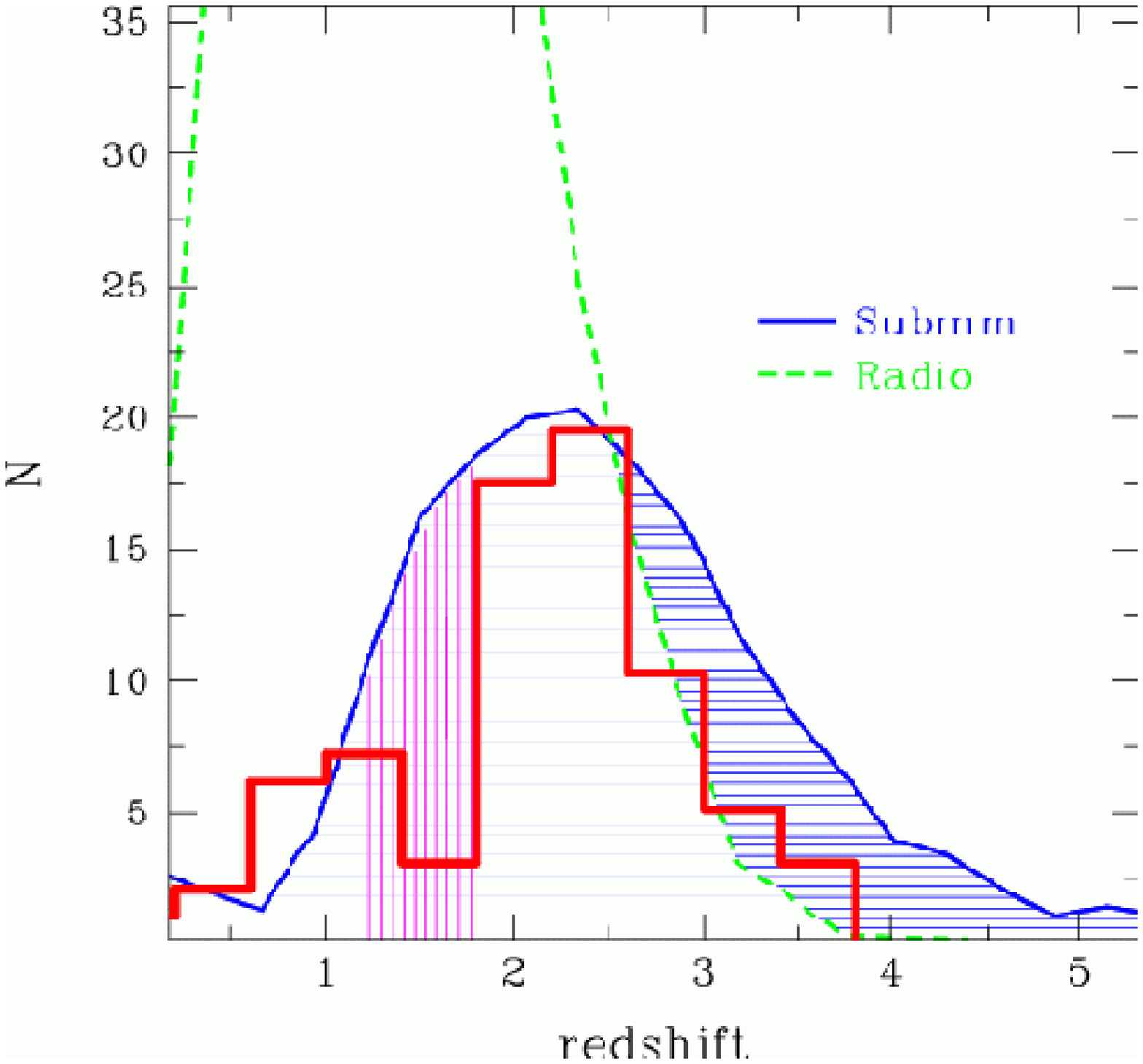}
\includegraphics[scale=0.37]{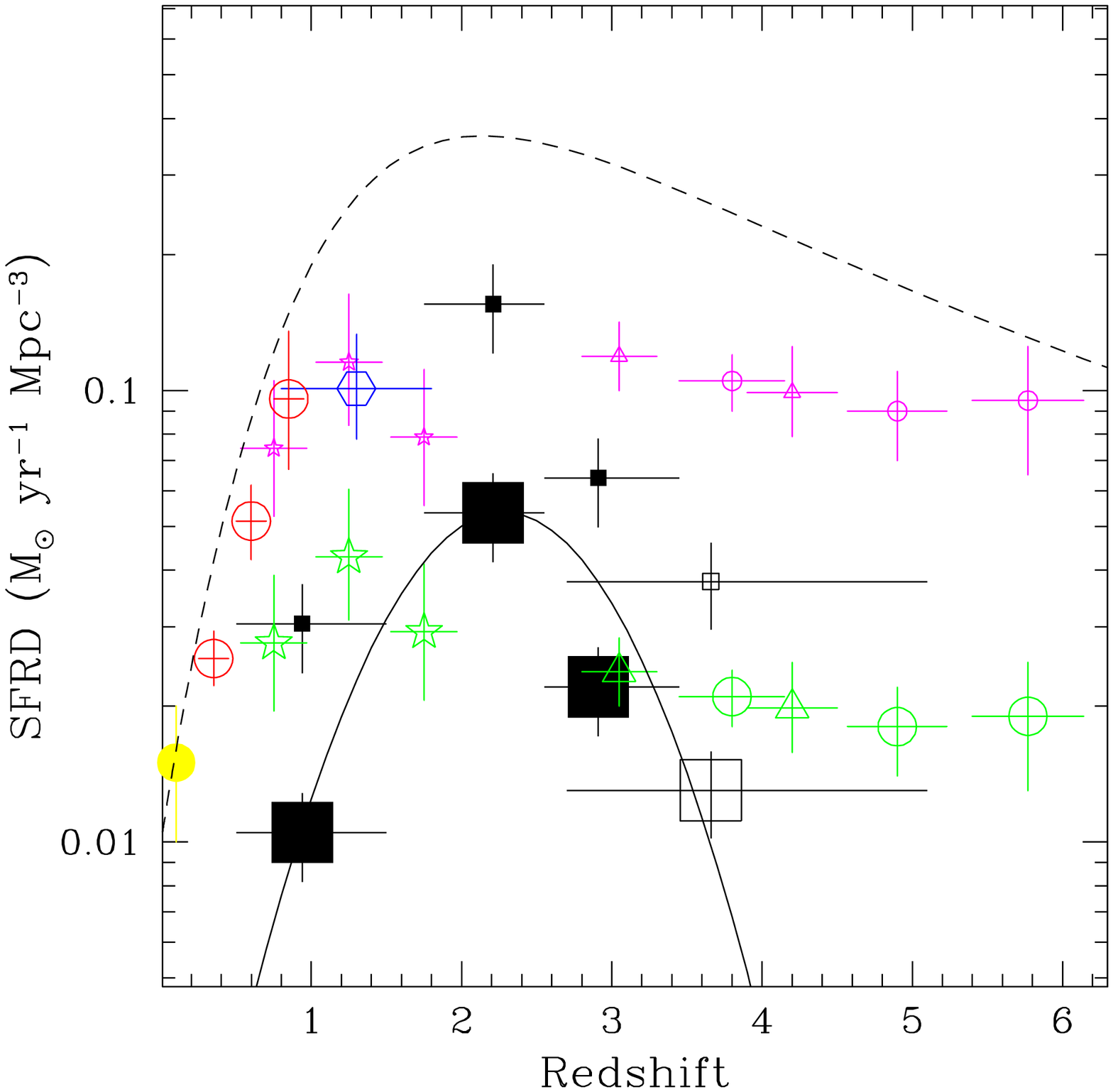}
}
\caption{
{\it Left:} Redshift distribution of the SMG sample spectroscopically identified by \cite{chapman05}
(red histogram). The blue line shows the model predicted redshift distribution, which may provide
on indication on the redshift bias due the radio selection criterion.
{\it Right:} Contribution of SMGs (large squares) to the evolution of the density of star formation
rate compared to estimates obtained by optical/UV surveys and radio/IR tracers (circles, triangles
and stars). Small squares show a correction to the luminosity density of SMGs to account for the
incompleteness due to the limiting flux of the redshift survey. See \cite{chapman05} for details.}
\label{fig_smg_z}
\end{figure}

\cite{chapman05} employed the alternative strategy of exploiting deep radio VLA observations.
Given the tight correlation between far-IR thermal emission
and radio synchrotron emission observed in galaxies, most SMGs should be associated with a detectable
radio source.
The higher angular resolution achievable with the VLA allowed \cite{chapman05} to locate
the position of the SMGs with an accuracy high enough to position the slit for the spectroscopic
identification.
Note also that the large VLA primary beam, resulting into a large field of view,
allowed the simultaneous radio detection of many SMGs in each single pointing.
The identification of a prominent Ly$\alpha$ in many of the deep spectra provided the
redshift for several tens SMGs.
The SMGs redshift distribution resulting from these surveys is shown in Fig.~\ref{fig_smg_z} (left), and
it is characterized by a median redshift $\rm \langle z\rangle \sim 2.3$. The redshift distribution
is very similar to that of QSOs, suggesting a link between the two populations of objects.

The sample of SMGs for which a redshift has been measured is limited to a submm flux of
a few mJy. The inferred redshifts imply far-IR luminosities of several times $\rm 10^{12}~L_{\odot}$,
i.e. extreme Ultra-Luminous Infrared Galaxies (more luminous, on average, than local ULIRG samples).
Therefore, currently identified SMGs represent only the most luminous population
of high-z dusty starbursts (the tip of the iceberg).
In Fig.~\ref{fig_smg_z} (right) the inferred contribution of SMGs to
the density of star formation (SFRD) is compared with those inferred by UV/optical surveys
and radio/IR tracers \citep{chapman05}.
SMGs are indicated with large squares, while small squares show an attempt to account for
the incompleteness due to the fact that current surveys only sample the most luminous sources,
as discussed above. Clearly, once incompleteness is accounted for, SMG contribute significantly
to the history of star formation, especially at redshifts around two. Again, their evolution
appears very similar to that of QSOs, suggesting a close link between SMGs and the evolution
of massive halos which hosts QSOs.

The redshift determination of SMGs allowed also their millimetric spectroscopic followup with
interferometers aimed at the detection of molecular transitions
(which requires an accurate knowledge of the redshift due to the limited
bandwidth). Intensive campaigns, especially with the IRAM PdB interferometer yielded the detection
of CO rotational transitions for several SMGs \citep{greve05}, indicating that these galaxies
also host huge amounts of molecular gas. The profile of the CO lines is often double-peaked,
indicative of a rotating disk or of a merging of two galaxies.
Followup interferometric observations at higher angular resolution have found that both the
continuum and line emission are very compact in most SMGs,
a few kpc or even less \citep{tacconi06,tacconi08}.
The compactness of the SMGs provides 
constraints on the inferred central densities, which result much higher than in any other population
of starburst galaxies, but comparable to those of compact passive galaxies
found in the same redshift range.
These findings, along with the
huge inferred star formation rates, suggest that SMGs are progenitors of local massive ellipticals,
experiencing their main episode of star formation, but which must undergo further structural
evolution in order to reach the sizes observed in local ellipticals.
These interferometric data also provide constraints on the global dynamical mass, typically yielding
masses of a few times $\rm 10^{11}~M_{\odot}$ \citep[e.g.][]{genzel03,genzel05}. Interestingly, the inferred
high density of massive galaxies ($\rm M>10^{11}~M_{\odot}$) at z$\sim$2
is difficult to reconcile with classical hierarchical models of galaxy evolution.

\subsection{Distant QSOs}
\label{sec_past_agn_highz}


Millimetric and submillimetric observations of distant AGNs are currently limited
mostly to powerful QSOs (although we will later discuss the case of lower luminosity
AGNs discovered in SMGs). Bolometric observations of large QSO samples have detected
mm-submm continuum emission at the level of a few mJy in about 60 optically selected QSOs
and about 20 radio galaxies at z$>$1, out to z=6.4 \citep{omont03,carilli01,priddey03,isaak02,
bertoldi03a,wang08}.
Generally the detection rate achieved with current facilities is about 30\% among bright QSOs.
Detected QSOs have far-IR luminosities of about $\rm L_{FIR}\sim 10^{13}~L_{\odot}$, but these
clearly represent the tip of the iceberg due to our current sensitivity limits.

\begin{figure}
\centerline{
\includegraphics[scale=0.65]{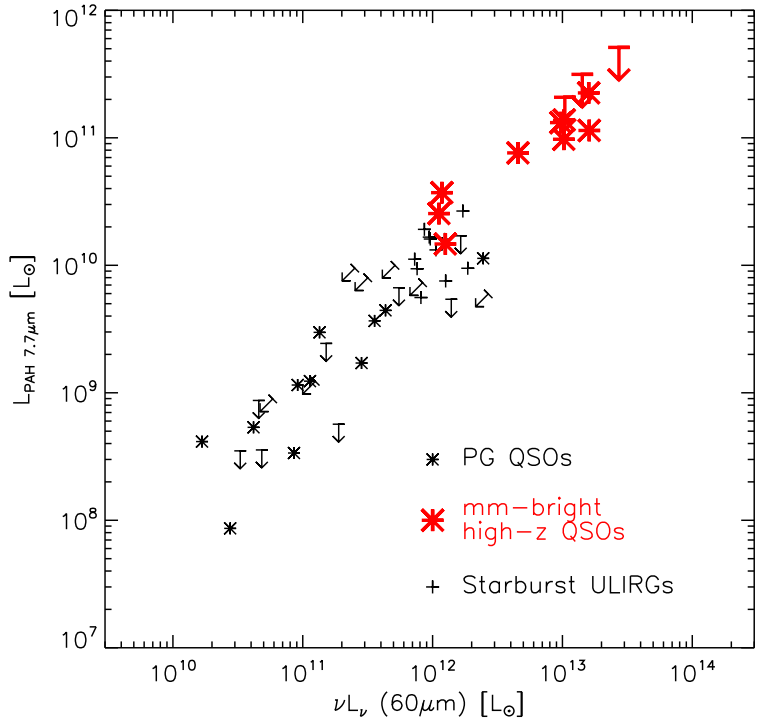}
}
\caption{
Relationship between PAH 7.7$\mu$m luminosity (which is a tracer of star formation) and far-IR luminosity
(inferred from the mm/submm emission in high-z QSOs). High-z mm-bright QSOs follow the same relation
of local starburst ULIRGs, indicating that the far-IR emission of the former is mostly due to star formation
\citep[from ][]{lutz08}.}
\label{fig_pah}
\end{figure}

For a fraction of high-z QSO the inferred far-IR luminosity follows the same correlation with
the radio luminosity as local star forming galaxies, suggesting that the far-IR luminosity is
powered mostly by star formation \citep{wang07}. This result is supported by the Spitzer
mid-IR detection, in a number of mm-bright QSOs, of strong PAH emission\footnote{The Polycyclic
Aromatic Hydrocarbon (PAH)
molecules are macro-molecules, or very small dust grains, which are excited by the (soft) UV radiation field
in star forming regions and emit strong features mostly at 6--11$\mu$m. These species are mostly
destroyed in the strong radiation field produced by AGNs.}, which is considered
a good tracer of star formation \citep{lutz07,lutz08}. Moreover, the PAH luminosity is found to correlate
with the far-IR luminosity and following the same relation of starburst galaxies (Fig.\ref{fig_pah}),
further supporting the starburst origin of the far-IR emission.

The star formation rates inferred from the far-IR luminosity (as well as from the PAH luminosity)
are as high as a few times $\rm 1000~M_{\odot}~yr^{-1}$. This suggests that in mm-bright QSO we are
witnessing the simultaneous growth of black holes (as traced by the optical and X-ray AGN emission)
and of the stellar mass in their host galaxy (as traced by the far-IR), which will probably evolve into
massive ellipticals \citep{granato04,dimatteo05}.
Such a co-coeval growth is expected in models of BH-galaxy evolution aimed at
explaining the local relation between spheroids and BH mass \citep{ferrarese00,geb00,marconi03}. 
However, the relation between star formation and black hole accretion (i.e. the relation between
$\rm L_{FIR}$ and $\rm L_{opt}$)
seems to saturate at high luminosities \citep{lutz08,maiolino07a}, suggesting that at high-z the BH growth
proceeds more rapidly than the host growth.


\begin{figure}
\centerline{
\includegraphics[scale=0.8]{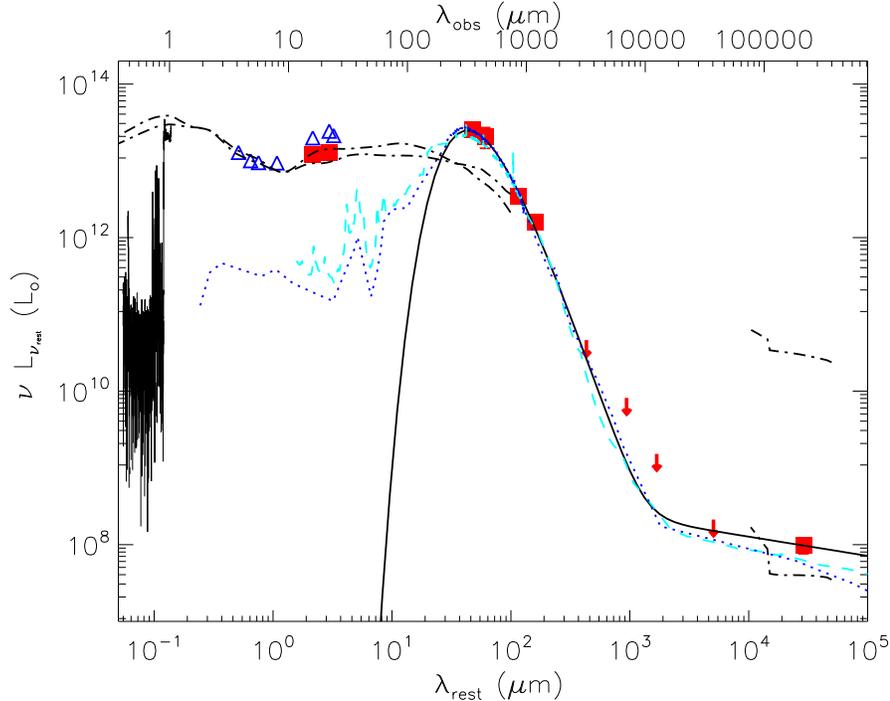}
}
\caption{
Near-IR to radio spectral energy distribution of one of the most distant QSOs,
J1148+5152, at z=6.4. Red and blue points show the observed photometry, while
curved lines show a combination of QSO and starburst templates fitting the data.
Note the prominent thermal far-IR emission, indicative 
of a large amount of dust (a few times $\rm 10^8 M_{\odot}$) already present
in this object close to the epoch of re-ionization.
Adapted from \cite{beelen06}.}
\label{fig_j1148_dust}
\end{figure}

The high far-IR luminosities inferred from the mm-submm observations also imply large amount
of dust in the host of high-z QSOs. Accurate measurements of the dust mass are however difficult to
obtain due to the unknown dust temperature and emissivity. Multiple band observations greatly help
to remove the degeneracy between these quantities \citep{beelen06,wang07}. Fig.\ref{fig_j1148_dust}
shows the rest-frame infrared SED of one of the most distant QSOs (z=6.4) known so far, for which
the far-IR thermal bump is relatively well sampled thanks to various submm
and mm observations. The inferred dust masses are as high as several times $\rm 10^8~M_{\odot}$.
The discovery of such huge masses of dust in the most distant QSOs is puzzling. Indeed, in the
local universe the main source of dust are the atmospheres of evolved stars (mostly AGB).
However, at z$>$6 the age of the Universe is less than 1~Gyr, which is the minimum time required
for AGB stars to evolve in large numbers and to significantly enrich the ISM with dust.
This suggests that in the early universe other mechanisms dominate the dust production.
Core-collapse SNe are candidate sources of dust on short timescales. Observations have indeed found
dust formed in SN ejecta \citep{sugerman06,rho08},
as also expected on theoretical grounds \citep{todini01,nozawa03}. The extinction curve inferred in high-z
QSOs and galaxies appears in agreement with that expected for dust produced in SN ejecta
\citep{maiolino04,stratta07,willott07}, supporting the idea that SNe may indeed be the major source
of dust in the early universe. However, even with the highest dust yield observed so far, the
huge mass of dust inferred from mm-submm observations of the most distant QSOs remains difficult
to account for \citep{dwek07}.


\begin{figure}
\centerline{
\includegraphics[scale=0.55]{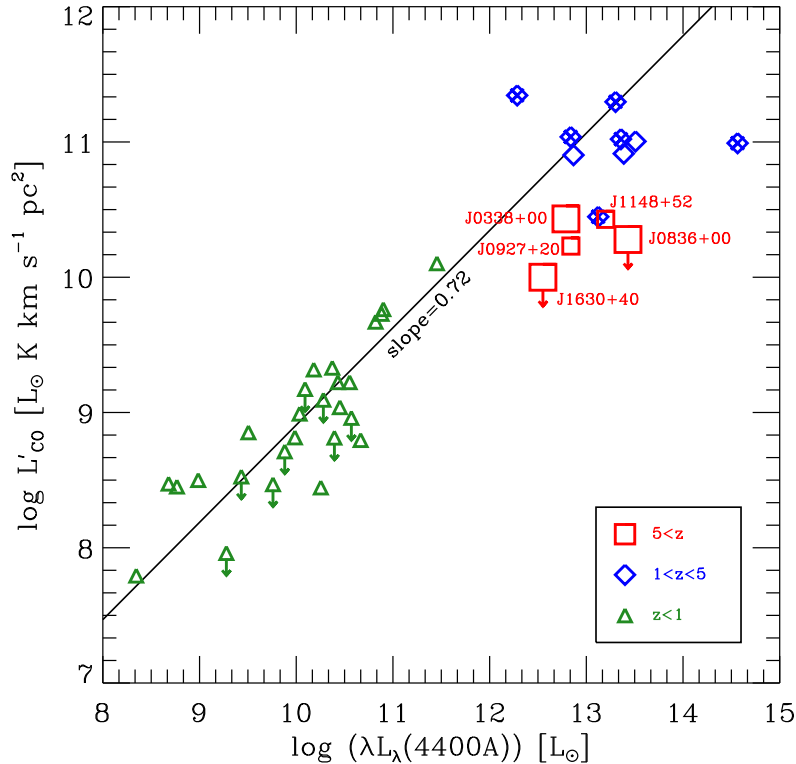}
}
\caption{
Correlation between the CO luminosity (tracing the molecular gas mass) and optical luminosity of QSOs
(tracing the black hole accretion rate). Note the non-linear relation between the two quantities.
From \cite{maiolino07b}.
}
\label{fig_co_opt}
\end{figure}

Extensive reviews on the CO emission in high-z objects, and in AGNs in particular,
are given in \cite{solomon05} and in \cite{omont07}.
More than half of the $\sim$40 CO detections obtained at high redshift (z$>$1) are in
AGNs (QSOs or radio galaxies). In particular, all of the detections at z$>$3.5 are in AGNs.
The higher detection rate in high-z QSOs is mostly due to their huge far-IR luminosities
and to the CO-FIR correlation (\S\ref{sec_past_normal_local}).
The molecular gas masses inferred from the CO detections are of the order of a few times
$10^{10}~M_{\odot}$. Except for the huge amounts of molecular gas, the general properties
of the CO emission in high-z QSOs and radio-galaxies
do not differ strongly from local and lower redshift powerful starbursts. The observation of
multiple CO transitions suggests
that the CO excitation temperature is higher in some QSOs \citep[e.g.][]{weiss07},
but the statistics are still very low. The
CO emission follows the same trend with the FIR luminosity observed in Fig.~\ref{fig_fir_co} , indicating
that QSO host galaxies are experiencing a strong starburst event in the process of rapidly exhausting
the available molecular gas, on a time scale of only $\rm \sim 10^7~yrs$. 
It is interesting to note that the relation between CO luminosity (tracing the amount of molecular
gas) and optical luminosity (tracing the black hole accretion rate in QSOs) is not linear, as
shown in Fig.~\ref{fig_co_opt}
\citep{maiolino07b}. This result indicates that, while the black hole can accrete
at very high rates (limited only by its Eddington luminosity), the host galaxy has only a limited
amount of molecular gas available for star formation (given by the galaxy evolutionary mechanism);
hence the two formation processes probably occur on different time scales (as already discussed above).

Deep observations have allowed the detection of transitions from other molecular and atomic species,
such as HCN, HCO$^+$, HNC, CN an CI, which are tracers of high density gas and of the gas chemistry
and excitation \citep[e.g.][]{carilli05,wagg05,riechers06,garcia06,guelin07,weiss05a}. However, these detections are limited to very few bright
sources, due to the limited sensitivity of current facilities. Within the limited statistics
available, the intensity of these lines relative to CO and to $\rm L_{FIR}$ do
not differ strongly from lower redshift and local starburst galaxies \citep[although there are indications
that the intensity of HCN decreases at high luminosities;][]{gao07,riechers07}.

\begin{figure}
\centerline{
\includegraphics[scale=0.55]{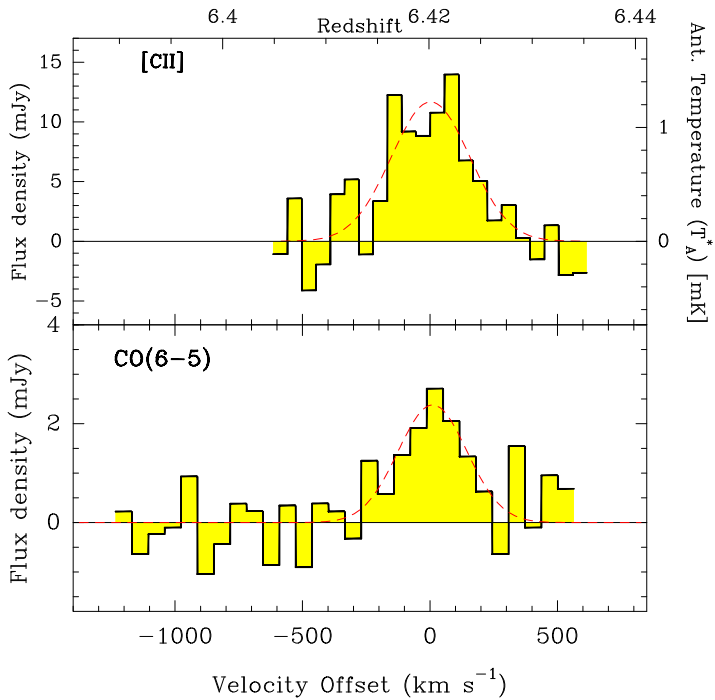}
}
\caption{
{\it Upper panel:}
Detection of the [CII]158$\mu$m fine structure line in one of the most distant quasars known, J1148+5152,
at z=6.4 \citep{maiolino05}. 
{\it Lower panel:} The CO(6--5) transition observed in the same objects is shown for comparison
\citep{bertoldi03b,walter03}. Note the different flux scales on the two panels.
}
\label{fig_cii}
\end{figure}

Recently, deep observations achieved the first detection of the [CII] fine structure line at 158$\mu$m
in two high-z QSOs (Fig.~\ref{fig_cii}) \citep{maiolino05,iono06}. This line is emitted by Photo-Dissociation
Regions (PDRs) in star forming galaxies, and it is the strongest line in the spectrum
of nearly any galaxy. The ratio $\rm L_{[CII]}/L_{FIR}$ observed in these two high-z QSOs is
about $\rm 2-4~10^{-4}$, i.e. similar to the value observed in local ULIRGs. The detection of [CII]
with a luminosity relative to $\rm L_{FIR}$
similar to local sources suggests that these high-z systems have already
been enriched with heavy elements, and with carbon in particular \citep{maiolino05}.


Generally the angular resolution and sensitivity are not good enough to trace CO
rotational curves and therefore to obtain accurate constraints on the dynamical mass.
However, in some high-z bright AGN the kinematics is resolved, especially in lensed cases, where the
lensing shear helps to resolve the galaxy structures \citep[e.g.][]{kneib98,venturini03,lewis02,
omont96,debreuck04,pap00}.
In other unresolved, or marginally resolved QSOs
the width of the CO line, along with measurements (or limits)
on the size of the CO emitting region provide some constraints on the dynamical mass \citep[e.g][]{maiolino07b}.
Typical dynamical masses inferred for QSO hosts and radio galaxies are of the order
of a few times $\rm 10^{11}~M_{\odot}$. However, we should recall that current
observations sample only the ``tip of the iceberg'' of the AGN population at high redshift.

Recently, \cite{ho07} attempted to calibrate the width of CO lines in galaxies as a proxy
of their dynamical masses. Once this calibration is applied
to the host of QSOs, the resulting galaxy mass can be compared with the black hole
mass inferred from their optical-UV emission lines. An interesting result is that high-z
QSOs appear to deviate from the local BH-galaxy mass relation, in the sense that their
hosts are less massive than expected. This result is in agreement with other independent
indications that the BH-galaxy mass relation evolves at high redshift \citep{peng06,mclure06}.
However, one should be careful when using the CO line width as a tracer of the dynamical
mass in (unobscured) QSOs. Indeed, selection effects make QSOs hosts to be viewed preferentially
face-on, since edge-on views tend to be obscured. As a consequence, the observed width of
the CO line in QSOs is likely reduced due to projection effects \citep{carilli06},
and may not be representative of the intrinsic rotation of the CO disk \citep{wu07}.

\begin{figure}
\centerline{
\includegraphics[scale=0.8]{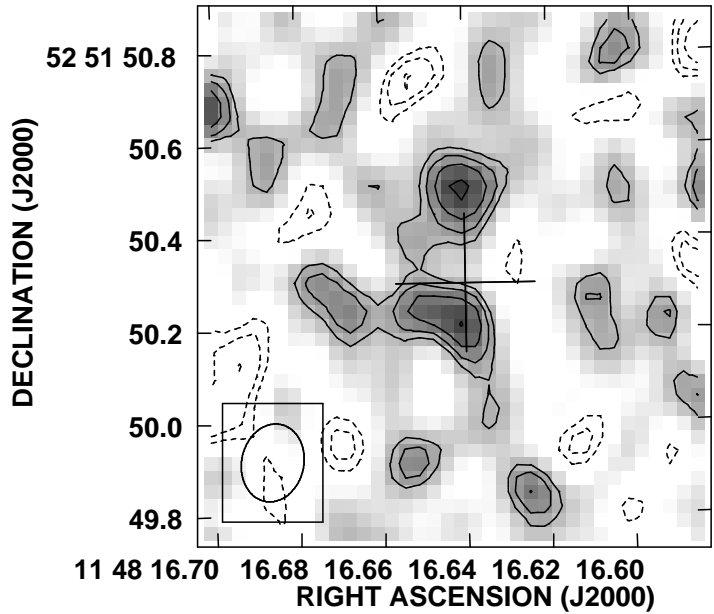}
}
\caption{
CO(3--2) map of the central region of J1148+5152, a QSO at z=6.4 \citep[from ][]{walter04}. At this redshift the angular
resolution of $\sim 0.15''$ corresponds to about 800~pc.
}
\label{fig_j1148_map}
\end{figure}

An interesting particular result is the CO(3--2) map of one of the most distant QSOs, J1148+5152,
at z=6.4 (Fig.~\ref{fig_j1148_map}) \citep{walter04}. The map shows resolved emission extending over about 5~kpc
and two central peaks separated by 1.7~kpc. From the velocity structure \cite{walter04} infer
a dynamical mass of about $\rm 5~10^{10}~M_{\odot}$. This mass is much lower than the
mass of a few times $\rm 10^{12}~M_{\odot}$ that would be expected from the black hole
mass derived for the same object (a few times $\rm 10^9~M_{\odot}$) if applying the local
BH-bulge mass relation. This result would provide further support to the scenario where at high-z
black holes grow faster then their host galaxies, while the latter
reach the local BH-bulge relation only at lower
redshifts. However, observations at higher angular resolution and with higher sensitivity
are certainly required to better determine the dynamical mass, and remove the degeneracy
with the inclination angle of the putative disk.


Above we have discussed mm-submm observations aimed at specifically
investigating objects known to host powerful AGNs (QSOs or radio-galaxies). We can however consider
the orthogonal approach: do galaxies discovered at mm-submm wavelengths (SMGs) host AGNs?
This issue has been investigated with the aid of deep X-ray observations. \cite{alexander03,alexander05}
found that most SMGs do host a X-ray source whose luminosity and spectral shape can only be ascribed
to the presence of an obscured AGN. However, even when corrected for absorption, the inferred
AGN luminosity cannot account for far-IR luminosity observed in SMGs, hence the main source
of far-IR emission in SMGs is due to star formation.
Similar results were obtained more recently thanks to Spitzer mid-IR spectroscopy \citep{valiante07,
menendez07,pope08}. The detection of strong PAHs in the spectra of most SMGs supports the scenario where
most of the luminosity is due to star formation. The detection of a weak mid-IR continuum due
to hot dust does trace the presence of an obscured AGN in some SMGs, but with a bolometric power
much lower than the starburst component.
Summarizing, SMGs represent a critical phase in the formation of massive galaxies; black hole
accretion is accompanying this major star formation event, but at a rate much lower than observed in QSOs.

\section{Current mm-submm facilities}
\label{sec_current_facilities}

Millimetric and submillimetric facilities can be divided in two main classes: single dish telescopes
and interferometers. These two classes of facilities are often complementary, especially in extragalactic
astronomy, one generally being characterized by wider field of view and higher sensitivity to continuum,
the other providing higher angular resolution and higher sensitivity for line detection.
This section provides a short overview of the current facilities in both classes,
by also discussing their main relative advantages and limitations.

\subsection{Single dish telescopes}
\label{sec_single_dishes}

A list of the main currently available single dish mm-submm telescopes is given
in Tab.~\ref{tab_sing_dish}. Current single dish telescopes have a diffraction--limited angular resolution
of only about 10--20$''$. However, one of the main advantages of single dish telescopes
relative to interferometers is their wide field of view, which makes single dish telescopes
optimally suited for wide area surveys. All single dish telescopes are equipped with various heterodyne
receivers for spectroscopy. However, an additional advantage of single dish telescopes is
the possibility of using bolometers, which (thanks to their wider band width)
are more sensitive to the continuum emission than heterodyne
receivers. To fully exploit the large field of view
for the detection of mm-submm sources, observatories have developed cameras hosting large bolometer arrays
that allow the simultaneous observation of wide areas of the sky. Tab.~\ref{tab_sing_dish} lists the main
bolometer cameras currently available, with a short summary of their characteristics (number of
detectors, wavelength of operation, field of view).
Most of the high redshift
submillimeter galaxies known so far have been discovered thanks to such bolometer arrays.
Some of these single dish observatories are dedicating an increasing fraction of their total observing time
to extensive surveys by exploiting these bolometer cameras. Probably this will be one of
the main working modes for some single dish telescopes when ALMA will be in operation: single dish
telescopes will provide continuum targets through wide area surveys to be then observed in depth with ALMA
(to obtain spectroscopic information, multi-band data, and angular resolution information).

\begin{table}
\caption{Main mm-submm single dish telescopes currently available}
\label{tab_sing_dish}
{\centering
\begin{tabular}{llllllll}
\hline\hline                 
Name & Diam. & $\Delta \lambda ^a$  & Beam ($\lambda$) $^b$ &
\multicolumn{4}{l}{Bolometer arrays} \\
     & [m]   &  [mm]             & [arcsec]  ([mm])              &
	 Name & N$_{det}^c$ & $\lambda ^d$ & FOV$^e$ \\
 & & & & & & [mm] & [arcmin$^2$] \\
\hline
IRAM & 30       &  1.2--3         &  11   (1.2) & MAMBO & 117 & 1.2 & 3$^f$  \\
JCMT & 15       &  0.45--1.2 &  15 (0.87) &
		SCUBA-2$^g$  & 10$^4$ & 0.45--0.87 & 55  \\
CSO  & 10.4     & 0.35--1.2   &  9 (0.35) &
		SHARC-II & 384 & 0.35--0.85 & 2.6  \\
	& & & &	Bolocam & 115 & 1.1--2.1 & 50$^f$  \\
APEX & 12       &  0.20--1.2 &  19 (0.87) &
		    LABOCA     & 295 & 0.87 & 102$^f$ \\
	& & & &	SABOCA$^g$ & 37 & 0.35 & 1.1$^f$ \\
ASTE & 10       &  0.35--0.87 & 17 (0.87) & \multicolumn{4}{c}{--} \\
Nobeyama & 45  &  3.4--10 & 15 (3) & \multicolumn{4}{c}{--} \\
LMT$^h$  &  50 & 0.85--4 & 7 (1.2) & AzTeC$^i$ & 144 & 1.1--2.1 & 2.4$^f$ \\
\hline\hline                 
\end{tabular}
}
\\
Notes:\\
$^a$ Wavelength range of operations (in units of mm).\\
$^b$ Beam size (FWHM), in arcsec, at the wavelength given in parenthesis
(in mm).\\
$^c$ Number of detectors in the array.\\
$^d$ Wavelength(s) of operations (in mm).\\
$^e$ Field of view of the array in arcmin$^2$.\\ 
$^f$ In these arrays the quoted FOV is not filled with detectors,
and the effective FOV (i.e. the region of the sky instantaneously sampled
in one single pointing) may be lower even by a factor of $\sim$4.\\
$^g$ Under commissioning at the time of writing.\\
$^h$ Under construction at the time of writing.\\
$^i$ While awaiting for LMT, AzTeC has already been successfully used at JCMT and ASTE.
\end{table}

It is worth noting that some of these single dish telescopes are also being equipped with very
broad band heterodyne receivers,
which will allow astronomers to identify the redshift of high-z sources through the
detection of multiple CO transitions
\citep[these instruments are often dubbed ``redshift machines'',][]{erickson07,
glenn07,stacey07,harris07}

However, one of the main problems of single dish telescopes in extragalactic astronomy is
their limited capability of detecting broad faint emission lines,
as a consequence of the (pseudo-continuum)
baseline variability. Although various techniques are employed to subtract the background
emission (position switching, beam switching, frequency switching), the background variability is often so
rapid and irregular that the resulting spectrum
is affected by a residual pseudo-continuum, generally tilted or curved. Such underlying curved
baseline is generally not much of a trouble for Galactic studies, where emission lines are very
narrow (hence the underlying baseline can be easily fitted and subtracted). However, this is
one of the main concerns for studies of faint objects in extragalactic studies, where lines are
relatively broad (a few 100~km/s) and may be confused or washed out by curved baselines.

\begin{table}
\caption{Main mm-submm interferometers currently available}
\label{tab_interf}
{\centering
\begin{tabular}{lcccc}
\hline\hline                 
Name & Antennas & $\Delta \lambda$  & Max ang. resol. & Total area \\
     &  [\# $\times$ Diameter]  & [mm]  &  [asec]     & [m$^2$]\\
\hline
IRAM-PdBI &  $\rm 6 \times 15m$ &  1.2--3         &  0.35 & 1060 \\
CARMA$^a$ & $\rm 6\times 10.4m+10\times 6m$ &  1.2--3   &  0.1  & 792  \\
NMA       &  $\rm 6\times 10m$ &  1.2--3         &  1    & 471 \\
SMA		  &  $\rm 8\times 6m$ &  0.35--1.2	   &  0.1  & 226 \\
eSMA$^b$    & $\rm SMA+15m+10.4m$   &  0.87--1.2   & 0.2 & 488 \\  
ATCA$^c$    & $\rm 6 \times 22m$ $^d$  & 3--12 &  2.  & 2280$^d$ \\
\hline\hline                 
\end{tabular}
}
\\
Notes:\\
$^a$ CARMA is the merging of the former OVRO (6$\times$10.4m antennas) and 
BIMA (10$\times$6m antennas) arrays.\\
$^b$ eSMA is the combination of SMA with the JCMT and CSO.\\
$^c$ ATCA can observe at much lower frequencies, down to $\rm \lambda = 20 cm$,
the specifications given here only refer to the observing
capabilities at mm wavelengths.\\
$^d$ At $\rm \lambda = 3 mm$ only 5 of the ATCA antennas can be used, for a total collecting
area of 1900~m$^2$.
\end{table}

\subsection{Interferometers}
\label{sec_interferometers}

Interferometers have obviously the advantage of much higher angular resolution relative
to single dish telescopes. Depending on their configuration and on the specific facility,
current interferometers yield
beam sizes ranging from a few arcsec to a few tenths of arcsec. Needless to say that
the most detailed maps of millimeter lines have been obtained with interferometers, both
in local galaxies and in distant systems.

Even when distant galaxies are not resolved,
interferometers remain the best tools to detect faint emission lines. The issue of
artificially curved/tilted baselines does not apply to interferometers, since
the source used for phase reference also calibrates and removes
the variations in sky transmission/emission. As a consequence,
faint emission lines in distant
galaxies were mostly detected with interferometers.

While superior for spectroscopy and line detection, interferometers have suffered limited
sensitivity to continuum emission. Indeed, until recently, interferometers have been equipped with
relatively narrow band receivers ($\Delta \nu \sim$500~MHz). Since the continuum sensitivity scales
as $\rm (\Delta \nu)^{1/2}$, this has made interferometers not competitive
for continuum detection relative to bolometers on single dish telescopes (whose band width is
generally limited by the adopted filters, $\Delta \nu \sim$50~GHz). However, more recently
some interferometers have been equipped with wider band receivers (e.g. the new IRAM-PdBI receivers
with $\Delta \nu \sim$4~GHz) therefore greatly improving their sensitivity to continuum.

Tab.~\ref{tab_interf} lists the main mm-submm interferometers currently available.
The IRAM array (located on the Plateau de Bure, French Alps) is the one with the largest collecting area,
and the most sensitive in the 1--3~mm range. CARMA (located on Cedar Flat, California)
is the recent merging of the
former BIMA and OVRO arrays; even with the combination of these antennas CARMA does not reach
the same collecting area of IRAM, but is expected to achieve higher angular resolution ($\sim 0.1''$).
SMA (located on Mauna Kea, Hawaii) is the only array working at submm wavelengths, down to 350$\mu$m. 
It can also work in combination with the JCMT and CSO (Tab.~\ref{tab_sing_dish}); this configuration,
dubbed as ``extended-SMA'' or ``eSMA'', doubles the total collecting area of the SMA, but it is limited
to $\lambda > 870 \mu$m.

\section{ALMA}
\label{sec_alma}

\begin{figure}
\centerline{
\includegraphics[scale=0.545]{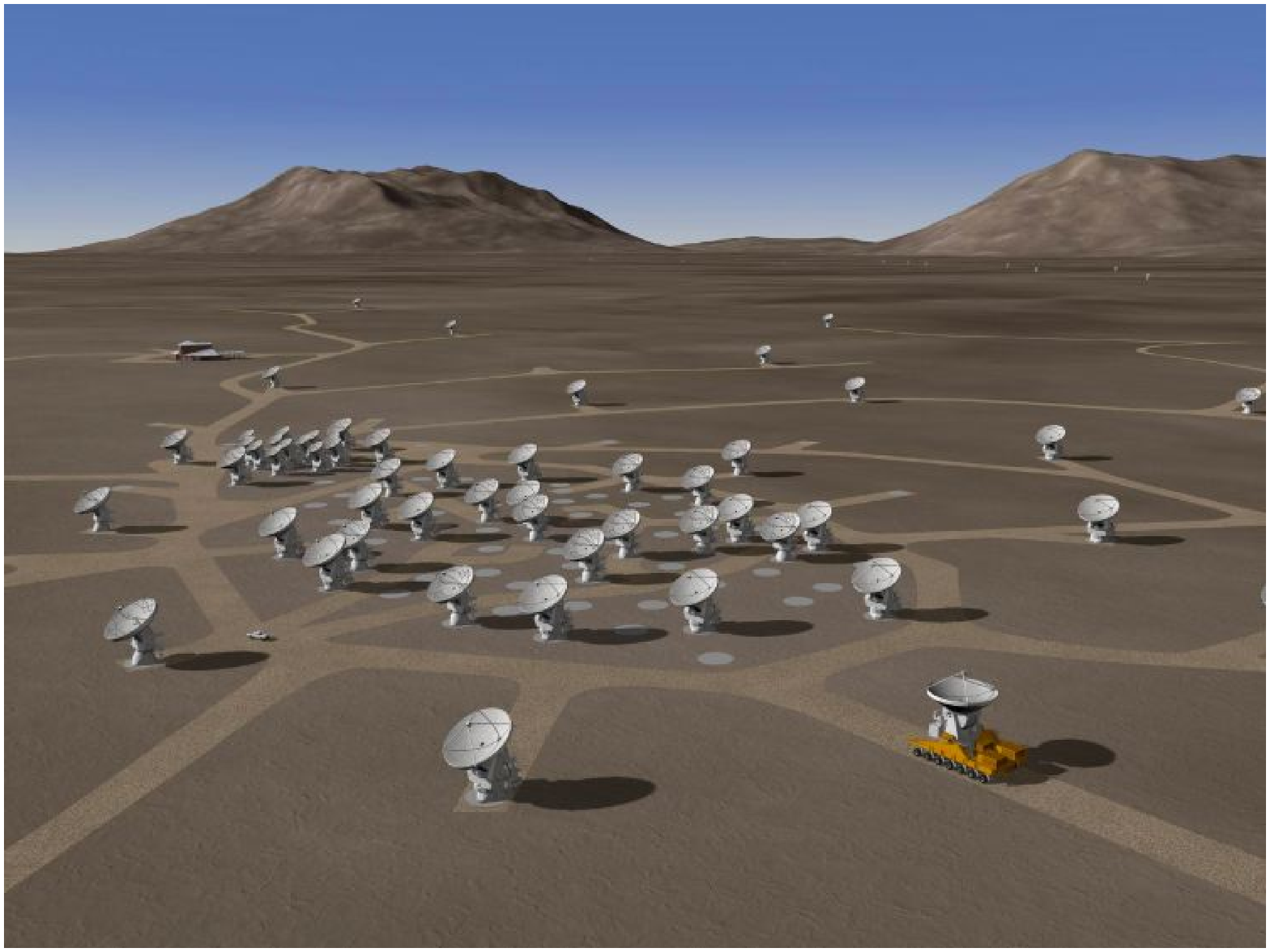}}
\vspace{0.1cm}
\centerline{
\includegraphics[scale=0.44]{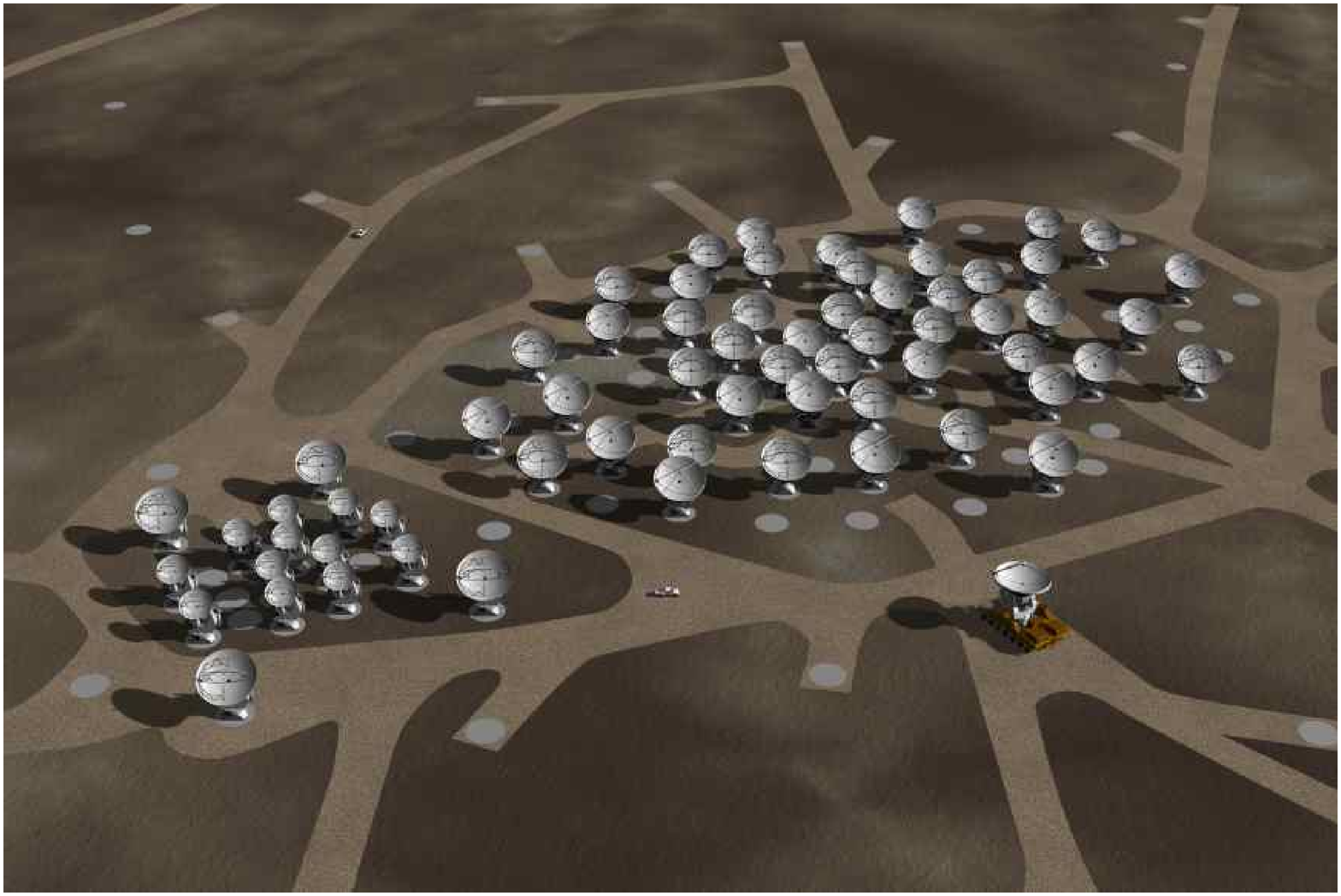}}
\caption{{\it Top:}
Artistic view of the ALMA array on Chajnantor Plateau, in an extended
configuration.
{\it Bottom:} Close up of the ALMA array in its compact configuration,
showing more clearly also
the Atacama Compact Array (ACA), located
in the bottom-left region.
The antenna transporter is also illustrated in operation in both panels
(bottom-right).
Courtesy of the European Southern Observatory.}
\label{fig_alma_art}
\end{figure}

The ``Atacama Large Millimeter Array'', ALMA, will provide a jump of nearly two orders of magnitude
in sensitivity and angular resolution with respect to existing facilities. It is therefore expected
to provide major step forwards and breakthroughs in many fields of astronomy.
Some specific key scientific cases, which drove the ALMA design, are:
1) The ability to detect spectral line emission from CO or C$^+$ in normal galaxies 
(Milky Way-like) at z$\sim$3 in less than 24~hours of observation.
2) The ability to image the gas kinematics in protostars and protoplanetary disks around young
Sun-like stars at a distance of 150~pc (i.e. roughly the distance of the star forming regions in
Ophiuchus or Corona Australis), enabling one to study their physical, chemical and magnetic field
structures and to detect the tidal gaps created by planets undergoing formation in the disks.
3) The ability to provide high quality images at an angular resolution of at least 0.1$''$.
The achievement of these scientific goals require: a) an array of antennas;
b) submm capabilities; c) a collecting
area $>6000~m^2$; d) a site which is high, dry, large and flat. These are the
considerations that mostly drove the design and development of ALMA.

The magnitude of the project is so large, both in terms of required funding and 
in terms of technical developments, that it necessarily needs the joint contribution of various
international institutes and agencies.
ALMA is now a world-wide project involving Europe
(through the European Southern Observatory, ESO), North America (U.S.A and Canada,
through the National Radio Astronomical Observatory, NRAO), East Asia (Japan and Taiwan,
through the National Astronomical Observatory of Japan, NAOJ), and the host country Chile.
Detailed information on the project can be
found at the web sites of these agencies (e.g. http://www.eso.org/projects/alma/,
http://www.alma.nrao.edu/,
http://www.nro.nao.ac.jp/alma/E/,
http://www.alma.cl/).
In the following I will only summarize the main technical features
of ALMA and its main observing capabilities.

\begin{table}
\caption{Frequency bands currently planned for ALMA}
\label{tab_alma}
{\centering
\begin{tabular}{lcccccc}
\hline\hline                 
Band & $\rm \Delta \nu$ & $\rm \Delta \lambda$ & ang. res.$^a$ & FOV$^b$ & cont. sens.$^c$ & line sens.$^d$\\
	& [GHz] & [mm] & [arcsec] & [arcsec] & [mJy] & [mJy] \\
\hline
3  &  84--116  &  2.6--3.6  &	3.0--0.034 & 56  & 0.05	 & 0.42 \\
4  & 125--169  &  1.8--2.4  &	2.1--0.023 & 48  & 0.06	 & 0.45	\\
5$^e$ & 163--211  &  1.4--1.8  &	1.6--0.018 & 35  & 0.77	 & 6.8	\\
6  & 211--275  &  1.1--1.4  &	1.3--0.014 & 27  & 0.10	 & 0.58	\\
7  & 275--373  &  0.8--1.1  &	1.0--0.011 & 18  & 0.20	 & 0.96	\\
8  & 385--500  &  0.6--0.8  &	0.7--0.008 & 12  & 0.37	 & 1.6 	\\
9  & 602--720  &  0.4--0.5  &	0.5--0.005 &  9  & 0.60	 & 2.1 	\\
10$^f$ & 787--950  &  0.3--0.4  &	0.4--0.004 &  7  & 0.90	 & 2.7	\\
\hline\hline                 
\end{tabular}
}
\\
Notes:\\
$^a$ Angular resolution in arcsec in the compact (200~m) and in the most extended configuration (18~km).\\
$^b$ Field of view in arcsec estimated as the Full Width Half Maximum of the primary beam. \\
$^c$ Continuum sensitivity: rms (in mJy) in one minute of integration, derived by using the ESO ETC. The
weather conditions were automatically selected by the ETC depending on the band.\\
$^d$ Emission line sensitivity: rms (in mJy) in one minute of integration, derived by using the ESO ETC
and by assuming a line width of 300~km/s.\\
$^e$ Initially only 6 antennas will be equipped with band 5 (which explains
the much lower sensitivity in this band).\\
$^f$ Funding of band 10 is still pending, awaiting results from the technical
development. The sensitivity and frequency range quoted for this band are just indicative, and may
be subject to significant variations depending on technical constraints.
\end{table}

ALMA consists of at least 54$\times$12m antennas and 12$\times$7m antennas for a total collecting area
of at least 6500~m$^2$. The antenna configurations will cover baselines ranging from 200~m out to 18~km,
yielding a maximum angular resolution better than 0.1$''$ at 3mm, and better than
0.01$''$ at submm wavelengths (Tab.~\ref{tab_alma}).
The field of view is limited by the primary beam of individual antennas and it ranges from 1$'$ at 3mm to
about 10$''$ at submm wavelengths (Tab.~\ref{tab_alma}).
Fig.~\ref{fig_alma_art} shows an artistic view of the ALMA array in extended and compact
configurations.
The 7m antennas and 4 of the 12m antennas are clustered into the
'Atacama Compact Array', ACA (Fig.~\ref{fig_alma_art}), which will have a compact fixed configuration aimed at
covering the {\it uv} plane on short spacings. ACA will be crucial to recover diffuse emission
extending on scales larger than a few arcsec (which is resolved out by the ALMA array).

\begin{figure}
\centerline{
\includegraphics*[scale=0.4]{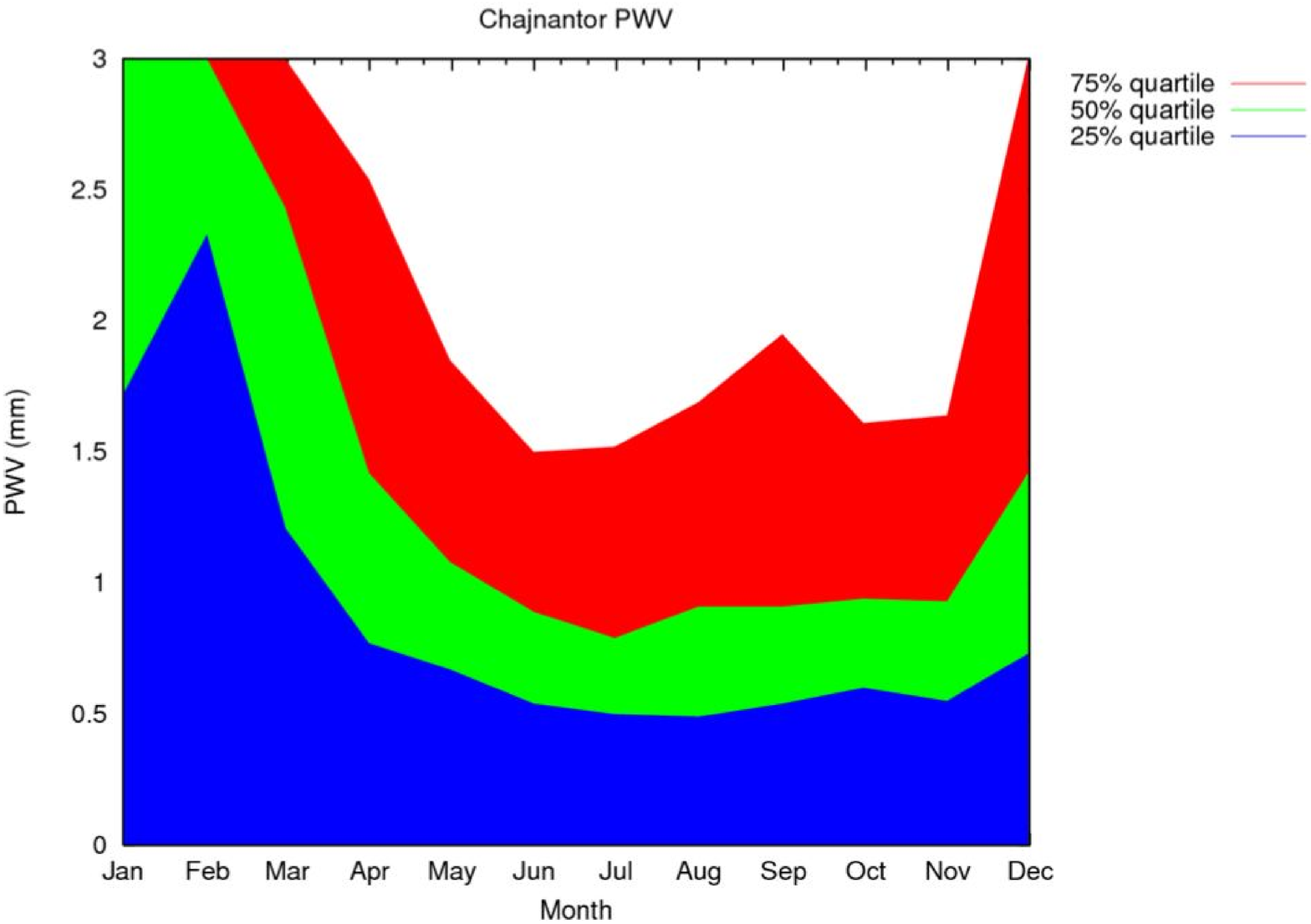}}
\caption{Statistics of the Precipitable Water Vapor (the main source of opacity at mm-submm wavelengths)
during the year at Chajnantor Plateau.
Note that during the winter months $\rm PWV < 1~mm$ for more than 50\% of the time,
allowing very sensitive observations at submm wavelengths. Courtesy of the European Southern Observatory.}
\label{fig_pwv}
\end{figure}

ALMA will be located on Chajnantor Plateau, in the Atacama desert (Chile), at an altitude of 5000~m.
This is an exceptionally dry site providing a good transparency even in the 350$\mu$m and 450$\mu$m
windows for a significant fraction of the time. In particular, the Precipitable Water Vapor (the main
source of opacity) is $\rm PWV < 1~mm$
for more than 50\% of the time, at least during the winter months (Fig.~\ref{fig_pwv}).
The vast flat topography of Chajnantor Plateau
is optimal to accommodate the various array configurations, providing some constraints on the location
of the antennas only in the most extended configuration. The antennas will be moved from one station
to the other by means of huge trucks at a rate of 2 antennas/day.

As listed in Tab.~\ref{tab_alma}, receivers for 
6 frequency bands are currently fully funded (bands 3, 4 and 6--9). 
Band 5 (163--211~GHz) is currently planned only for 6 antennas. The 
funding and the specifications of band 10 are pending,
since the receivers development is still in progress.
The distribution of the ALMA bands relative to the atmospheric windows is illustrated in
Fig.~\ref{fig_atm_transmission}.
The maximum simultaneous frequency coverage allowed by the huge correlator is 8~GHz, which allows
ALMA to reach high sensitivities even for the continuum detection. Tab.~\ref{tab_alma} summarizes the
ALMA sensitivities (both for continuum and line detection). For instance, in band 7 (870$\mu$m)
ALMA reaches the extraordinary sensitivity of $\sim$0.1 mJy (rms) in one minute of integration,
to be compared with the SCUBA sensitivity of about 1 mJy (rms) in 1 hour of integration.

Constructions at the ALMA site started already a few years ago and are now nearly completed.
At the time of writing the
first eight antennas have arrived for integration and testing at the Operations Support Facilities
(OSF, located at about 30~km from the ALMA site). In early 2009 the first three antennas will be moved
to the ALMA Operations Site (AOS) for commissioning. In 2010 a call for proposal is planned
for early science with an initial set of at least
16 antennas. The full array is scheduled to come in full operation
in 2012.

An important feature of the ALMA organization is the plan of making it an observing facility easily accessible
also by non-experts in the field. Both the software to prepare the observations and the
pipeline for the data reduction are designed to be easy to use also by people that are not necessarily
experts in interferometry nor necessarily acquainted with mm/submm observing techniques.
This is an important characteristic that will make ALMA totally open to a broad astronomical
community.

\section{ALMA prospects for AGN studies}
\label{sec_alma_prospects}

It is obvious that the ALMA capabilities will allow astronomers to tackle numerous outstanding issues,
spanning most of the research fields in astronomy. A summary of some of the main scientific goals can
be found in the Design Reference Science Plan (DRSP), available at
 http://www.strw.leidenuniv.nl/∼alma/drsp11.shtml.
It is beyond the scope of these lecture notes to provide an overview of the several
outstanding scientific cases for ALMA.
In this section I will only focus on some of the ALMA prospects for AGN studies and some
related topics, but even in this field the discussion is not meant to
be exhaustive.

\subsection{Local AGNs}
\label{sec_alma_local}

\begin{figure}
\centerline{
\includegraphics[scale=0.6]{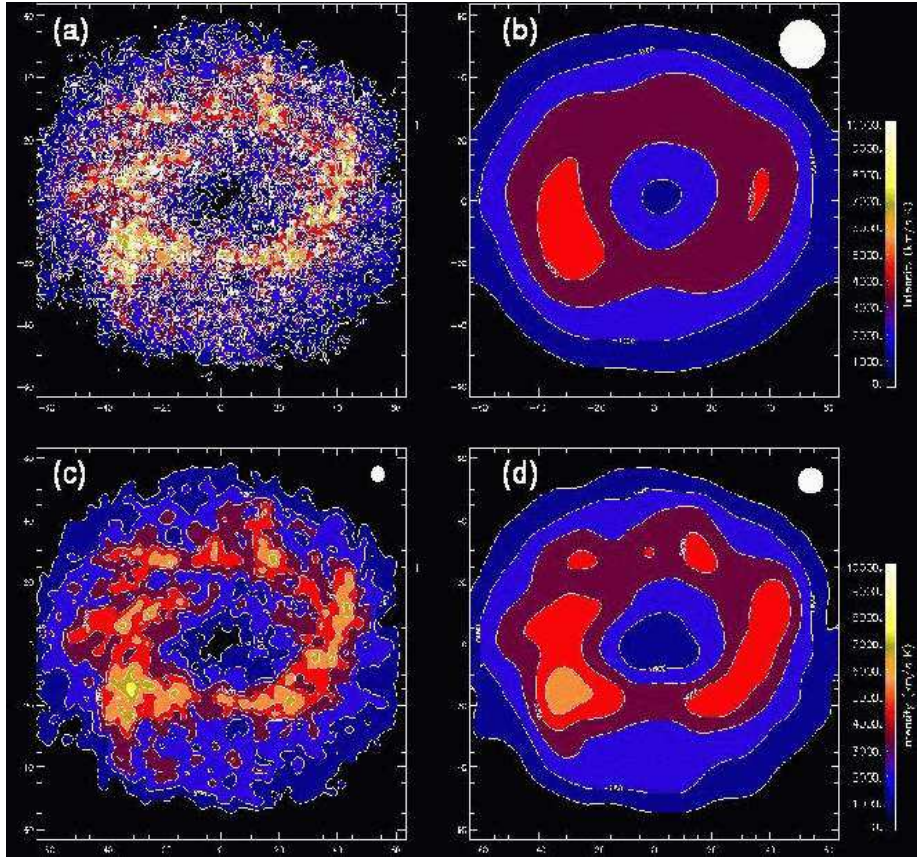}}
\caption{CO(2--1) intensity map of NGC~1068 resulting from the extended torus AGN model in \cite{wada05}.
The top-left
panel shows the original map obtained by the model,
the other panels show the expected morphology after convolution with different
beam sizes. The map expected at the ALMA resolution is shown in the bottom-left panel (c).
In this model, the physical size of the torus is about 64~pc in diameter.}
\label{fig_wada}
\end{figure}

One of the most hotly debated issues on AGNs is the
physics, structure and dynamics of the circumnuclear molecular medium, which is both responsible for
the obscuration of AGNs
and presumably related to their feeding. Some models assume a uniform gas distribution
in a toroidal geometry, whose thickness is supported by IR radiation pressure \citep[e.g.][]{krolik07}.
Other models propose that the circumnuclear obscuring medium is clumpy and originated by the
outflow of the accretion disk \citep{elitzur06,nenkova02,honig06}.
Yet other models assume a clumpy distribution due to
the effects of supernova explosions or stellar winds, which are also responsible for making
the medium ``thick''
\citep{wada05,watabe05,naya07}.
The high angular resolution, along with its spectroscopic capabilities, will allow ALMA to clearly
distinguish between these models, both through morphological analyses and by investigating the
kinematics of the nuclear and circumnuclear medium. Fig.~\ref{fig_wada} shows the expected distribution of
circumnuclear CO emission according to the torus model in \cite{wada05}, and convolved with the
angular resolution of ALMA projected at the distance of NGC~1068 (panel {\it c}): the clumpy nature
of the ``torus'' is clearly resolved.

Actually, the structure shown in Fig.~\ref{fig_wada}
describes the extended
component of the torus (radius $\sim$32~pc), which contains a large amount of gas, but accounts
only for a small fraction of the nuclear obscuration. Most of the obscuration is due to dense gas
on the parsec scale \citep{maiolinorisaliti07}, as also inferred by mid-IR interferometric observations
\citep{jaffe04}. In clumpy models the dust within each individual cloud spans a wide range of
temperatures and therefore emits strongly at all wavelengths, from mid-IR to submm. As a consequence,
an interesting prediction of clumpy models is that the nuclear ``torus'', made of out of several
clouds, should show the same morphology at all infrared wavelengths\footnote{Actually some difference
of the morphology as a function of wavelength is expected, but well predicted by models.}.
In contrast, uniform ``torus'' models expect a strong radial temperature gradient and, therefore, the
innermost warm dust should emit much more at mid-IR wavelengths, while submm radiation from cold
dust should be observed mostly in the outer regions. At high frequencies ALMA has an angular resolution
matching VLTI (i.e. sub-pc scale at the distance of famous AGNs such as NGC~1068 and Circinus),
therefore it will be possible to directly compare the mid-IR morphology with the submm morphology
and directly test the predictions of clumpy and uniform torus models.

The gas kinematics inferred from ALMA observations of the nuclear molecular emission lines will provide
important information on the dynamical stability and origin of the circumnuclear medium. In the case of
a simple uniform structure, supported by radiation pressure, we expect gas motions to be dominated
mostly by rotation. In the case of a medium originated by the accretion disk wind \citep{elvis00}
we expect the molecular gas kinematics to also show a clear outflow component.
In the case of a strong contribution
by SN and stellar winds we also expect a strong turbulent component, hence a large width of the
emission lines even within resolved regions.
Obviously investigating the molecular gas in the innermost regions will require the observation
of species typically tracing dense and highly excited gas (e.g. HCN, HCO$^+$,...), which are
much fainter than CO, but easy to detect with ALMA. A detailed modelling of the excitation and
emissivity distribution of these high density tracers in AGN torii has been recently performed by
\cite{yamada07}. According to these models the flux ratios between these
molecular lines is expected to be
spatially highly inhomogeneous, reflecting the inhomogeneous structure of the molecular
torus. Only the excellent angular 
resolution of ALMA will allow us to reveal such complex structures for the first time.

\begin{figure}
\centerline{
\includegraphics[scale=0.7,bb=65 200 580 635]{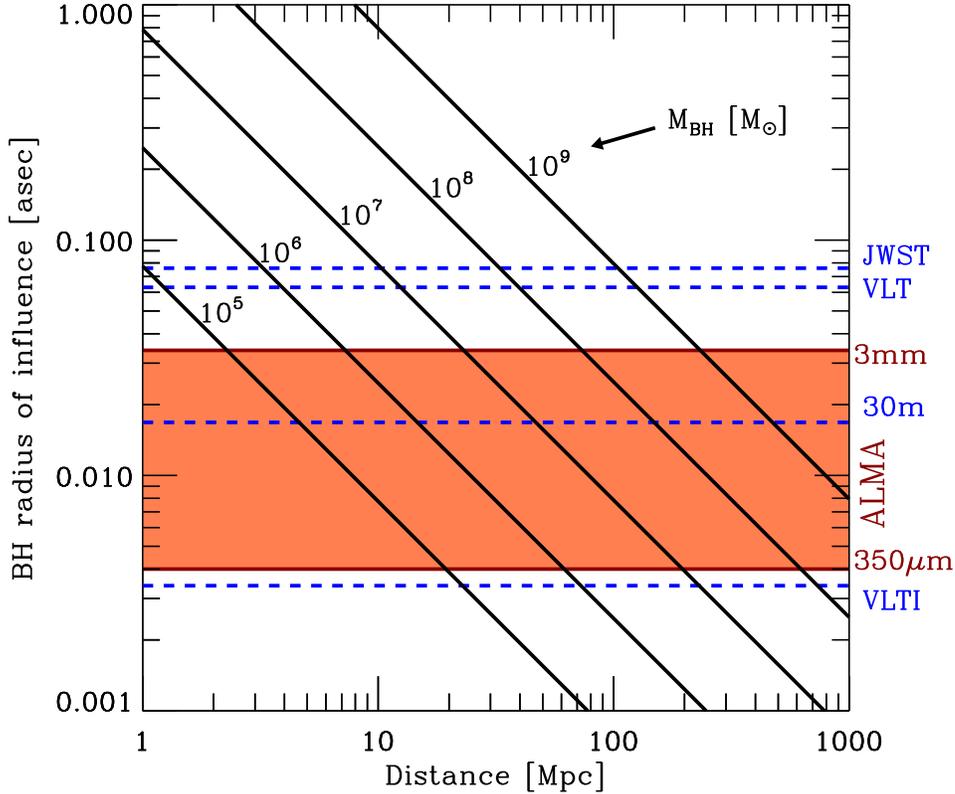}}
\caption{Radius of influence (projected on the sky) of black holes of different masses as a function
of the distance of their host galaxy.
The radius of influence was calculated by assuming a stellar gravitational
potential matching the $\rm M_{BH}-\sigma$ relation. The horizontal blue dashed lines indicate
the angular resolution at $\rm \lambda = 2\mu m$ of various current or planned optical/near-IR
facilities (JWST, 8m telescope, 30m telescope, VLTI).
The red shaded area gives the
range of angular resolution achievable with ALMA at different frequencies. Courtesy of A. Marconi.}
\label{fig_bh_Rinf}
\end{figure}

By resolving the molecular gas dynamics on such small scales it will be possible to measure the
mass of the nuclear black hole. For instance, in NGC~1068 the $\rm 10^7~M_{\odot}$ black hole
has a sphere of influence with a radius of about 4~pc, requiring a beam projected
on the sky smaller than 0.05$''$ to be resolved, which is achievable with ALMA at
most frequencies. In NGC~1068, and a few additional nearby Sy2s, the BH mass has been
already measured thanks to VLBI observations of a maser H$_2$O disk \citep{greenhill96}; however, the latter
technique can only be exploited in a few cases where the maser disk is observed nearly edge-on
(within 15$^{\circ}$ of the line of sight). ALMA observations of the nuclear
molecular gas will allow us to extend the measurement of black hole masses in AGNs to large
samples, without stringent constraints on the geometry of the nuclear gas distribution.
Fig.~\ref{fig_bh_Rinf} shows the radius of influence, as projected on the sky, of black holes of different
masses in the nuclei of galaxies, as a function of their distance (the central stellar gravitational
potential is assumed to match the $\rm M_{BH}-\sigma$ relation). The red shaded
area indicates the range of angular resolution achievable with ALMA at various frequencies. Clearly
ALMA is in principle able to measure even very small black holes ($\rm M_{BH}<10^5~M_{\odot}$)
in nearby systems ($\rm D<10~Mpc$); therefore ALMA will allow us to extend the investigation
of the  $\rm M_{BH}-\sigma$ relation to very low masses. ALMA will also allow us to measure
massive black holes ($\rm M_{BH}>10^7~M_{\odot}$) in distant systems, $\rm D\sim
50-1000~Mpc$; at these distances we can find the closest cases of merging systems
(most of which are powerful ULIRGs). In these cases ALMA will measure
the black hole mass within each of the two merging systems, which will allow us to place the two
individual black holes on the
$\rm M_{BH}-\sigma$ relation and to compare their location with the evolutionary path expected
by models of hierarchical black hole growth \citep[e.g.]{dotti07,volonteri03}.

\begin{figure}
\centerline{
\includegraphics[scale=0.63]{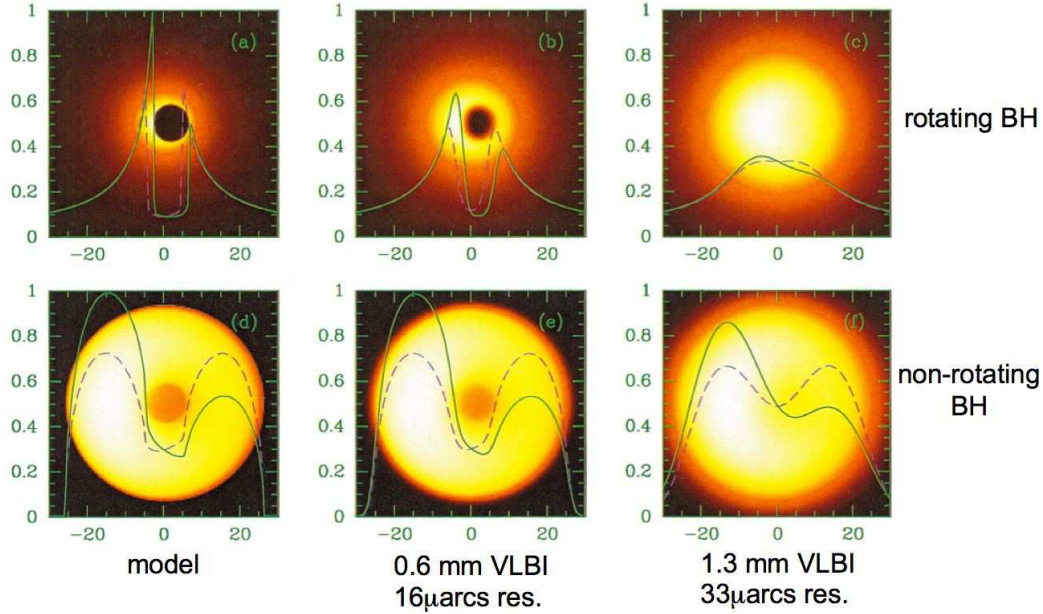}}
\caption{Expected images of SgrA$^*$ at mm-submm wavelengths with the angular resolution achievable with
a mm-submm VLBI. The left panels illustrate the result of the model, showing a clear ``shadow'' due to the
light bending by the gravitational field of the supermassive black hole. The other panels show the
image obtained after convolution with the beam expected for the mm-submm VLBI. The panels on the top are
for a rotating black hole, while the panels on the bottom are for a non-rotating black hole.
From \cite{falcke00}.
}
\label{fig_falcke}
\end{figure}

ALMA will also be a powerful machine to discover heavily obscured AGNs.
It is now clear that a population of obscured AGN are missed
by optical spectroscopic surveys aimed at identifying AGNs through their characteristic
narrow emission lines \citep{imanishi07,franceschini03,ballo04,caccianiga07,
maiolino03}. Some of these obscured AGNs are missed because
the Narrow Line Region is also obscured by dust in the host
galaxy \citep{haas05}. In other cases
the NLR is not formed at all \citep[as inferred by the lack of NLR lines even in the mid-IR, ][]{armus07},
probably because the ionizing radiation is absorbed
in all directions (i.e. 4$\pi$ obscuration rather than a torus-like geometry).
However, the X-rays emitted by the AGNs create extended XDRs, which are characterized by enhanced
temperatures favoring the formation of several molecular species characteristics of these
regions, as discussed in \S\ref{sec_past_normal_local}.
The molecular transitions and diagnostics typical of these
regions \citep[e.g. enhanced HCN/HCO$^+$][]{gracia06,kohno08}
do not suffer any dust absorption and will be easily detected with
ALMA even in distant systems. Therefore, ALMA will be able to provide an unbiased census
of the AGN population, by including also those heavily obscured AGNs that are not detectable 
at other wavelengths.

I conclude this section by discussing the potentialities for the investigation
of our Galactic center. The nuclear radio source SgrA$^*$ is also a mm/submm source.
If the emitting region is uniformly distributed around the supermassive black hole, then
\cite{falcke00} showed that photons are expected to be deviated by the strong gravitational field
within a few Schwarzschild radii, therefore creating a ``shadow'' in a putative mm/submm high resolution
image (Fig.~\ref{fig_falcke}, left).
There is a plan of combining ALMA with other mm/submm observatories distributed world wide to create 
(sub-)millimetric VLBI network, reaching an angular resolution of a few 10$\mu$arcsec, which would
allow us to resolve the black hole ``shadow'' (Fig.~\ref{fig_falcke}, right). This would really be a major result,
essentially the first ``picture'' of a black hole. The shape and the contrast of the ``shadow''
would also allow us to determine whether the black hole is rotating (upper panels in Fig.~\ref{fig_falcke})
or not (lower panels).

\subsection{High redshift galaxies and AGNs}
\label{sec_alma_highz}

We have seen in \S\ref{sec_past_starburst_highz}
that current mm/submm observations of high-z objects are limited
to very luminous systems ($\rm L_{FIR}\sim 10^{13}~L_{\odot}$),
not representative of the bulk of the galaxy population
at high redshift. ALMA will be able to detect galaxies at least two orders of magnitude
fainter, therefore providing an unbiased view of the evolution of galaxies from the
mm/submm perspective. Fig.~\ref{fig_sed_z}, compares the continuum sensitivity of ALMA (24 hours of integration)
 with the continuum flux expected by a galaxy with $\rm L_{FIR}=10^{11}~L_{\odot}$ (i.e. $\rm SFR\sim
 15~M_{\odot}~yr^{-1}$) at various redshifts. It is clear that with the same integration time
ALMA will be able to detect the thermal dust continuum of galaxies that are three times fainter.
It is also interesting
to note that for a SED typical of starburst galaxies the ALMA sensitivity nicely matches
the mid/near-IR sensitivity of JWST,
another major facility which will be launched in 2013, i.e. when ALMA will
be already fully operating. These two facilities will be fully complementary for the investigation
of faint distant galaxies that are out of reach for current observatories.

\begin{figure}
\centerline{
\includegraphics[scale=0.85,bb=90 250 540 585]{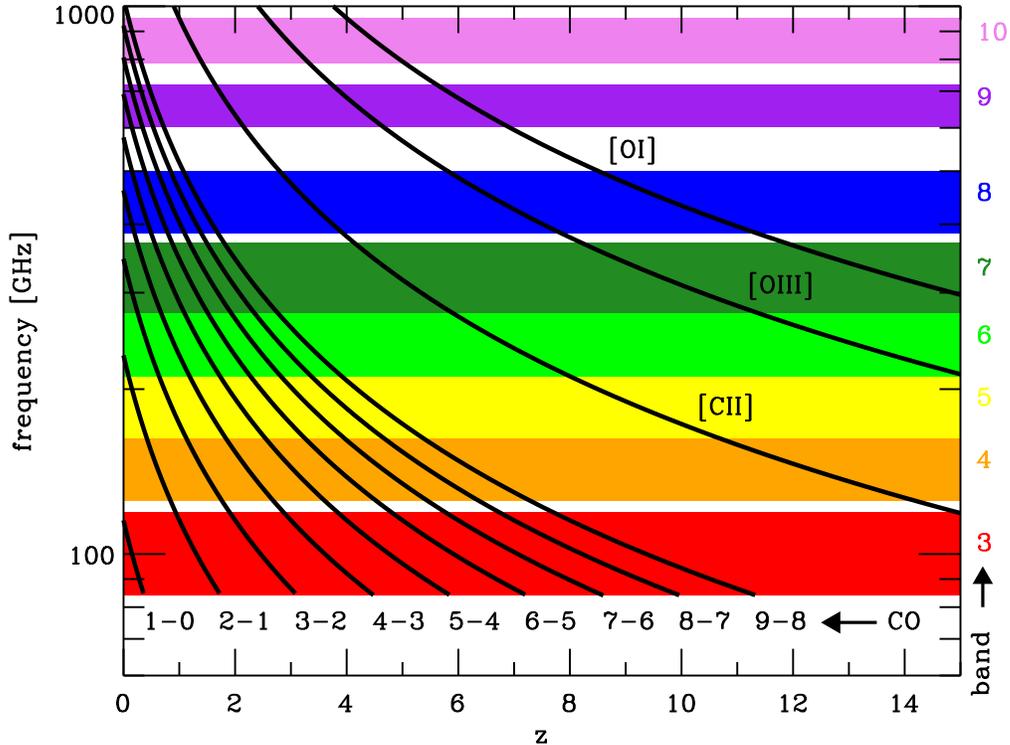}}
\caption{Frequency of the various CO rotational transitions and of the [CII]158$\mu$m, [OI]63$\mu$m
and [OIII]88$\mu$m fine structure lines as a function of redshift.
The colored shaded bars indicate the frequency of
the bands currently planned for ALMA.}
\label{fig_lines_z}
\end{figure}

\begin{figure}
\centerline{
\includegraphics[scale=0.9,bb=120 260 495 565]{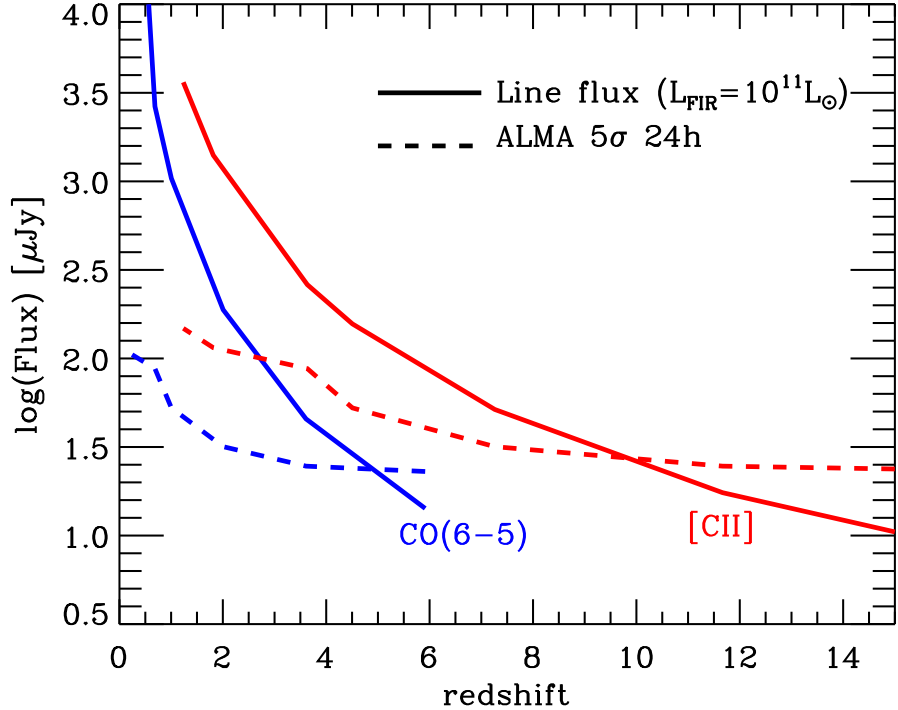}}
\caption{{\it Solid lines:} CO(6--5) and [CII]158$\mu$m line fluxes expected
for a galaxy with $\rm L_{FIR}=10^{11}~L_{\odot}$ as a function of redshift (keep in mind that the
luminosity of these lines does not scale linearly with $\rm L_{FIR}$). {\it Dashed lines:}
Sensitivity of ALMA (24 hours of integration)
for a 5$\sigma$ detection of emission lines (width of 300~km/s) in the frequency bands corresponding
to the CO(6--5) and [CII]
redshifted lines. Here I have only considered the frequencies around the center of the bands and
far from deep atmospheric absorption features (Tab.~\ref{tab_alma}), and I have neglected
band 5 (which has currently
much lower sensitivity because available only for 6 antennas). The [CII] luminosity was inferred by
fitting to the $\rm L_{[CII]}/L_{FIR}$ relation reported in \cite{maiolino05}.
The CO luminosity was inferred by assuming the $\rm L_{C0}-L_{FIR}$ relation in Fig.\ref{fig_fir_co} and
by assuming the relative luminosities of the CO transitions observed in SMM J16359+6612 (z=2.5)
by \cite{weiss05b}.}
\label{fig_alma_lines}
\end{figure}

Even in its compact configuration ALMA will have a resolution high enough ($\sim 1''$) to make
the confusion noise negligible, which is instead one of the main limitations of current surveys
with single dish telescopes. The excellent ALMA resolution will allow astronomers to identify
the optical/near-IR counterpart of mm-submm sources without ambiguities.
The optical/near-IR counterpart may provide the redshift of the galaxy/AGN through optical/near-IR
followup spectroscopy. However, ALMA will directly provide the redshift of the mm/submm
sources through the detection of molecular lines or fine structure atomic lines. Fig.~\ref{fig_lines_z} shows the
observed frequency of the CO rotational transitions as a function of redshift overlaid on the
ALMA frequency bands. At z$>$3 at least two CO transitions are observable within band 3 (the most sensitive one),
therefore unambiguously providing the redshift of the source. At z$<$3 the redshift confirmation  requires
the observation an additional CO line in a higher frequency band.
Fig.~\ref{fig_alma_lines} illustrates the ALMA sensitivity for the detection of the CO(6--5)
line \citep[which in SMGs is close to the peak of the CO lines intensities, ][]{weiss05b} for
a galaxy with $\rm L_{FIR}=10^{11}~L_{\odot}$. ALMA will clearly be able to detect this CO transition
in Luminous Infrared Galaxies (LIRGs) out to z$\sim$5. Note that to scale the diagram
in Fig.~\ref{fig_alma_lines} to other FIR luminosities one has to keep in mind the non-linear
relation between $\rm L_{CO}$ and $\rm L_{FIR}$ (Fig.~\ref{fig_fir_co}).

At z$>$7 only CO transitions higher than (6--5) will be observable within the ALMA bands. However, such
high transitions are generally little excited in most galaxies \citep{weiss05b} and therefore more
difficult to observe relative to the lower transitions. Yet, already at z$>$1
the strong [CII]158$\mu$m line will be observable in the ALMA bands,
and it will be one of the main tools to identify the redshift of distant sources, and in particular
at z$>$7 (Fig.~\ref{fig_lines_z}).
Fig.~\ref{fig_alma_lines} shows that in LIRGs ($\rm L_{FIR}>10^{11}~L_{\odot}$) ALMA will be able
to detect [CII] out to z$\sim$10. Note that also in this case the diagram cannot be scaled
linearly to other FIR luminosities, since the relation between $\rm L_{[CII]}$ and
$\rm L_{FIR}$ is not linear \citep{luhman03}.

The second brightest line is generally [OI]63$\mu$m (in some objects 
this line is even stronger than [CII], but in some exceptional objects it may be self-absorbed).
[OI]63$\mu$m will be observable at z$>$4
in the ALMA bands, and will help to obtain the redshift confirmation along with the [CII] detection.
The [OIII]88$\mu$m line is also generally relatively strong, especially if an AGN is present
($F_{[OIII]}\sim 0.5-0.3 ~F_{[CII]}$), and can be detected with ALMA at high redshifts. However, since
[OI] and [OIII] will be observable at high-z mostly in the high frequency bands, which are the least
sensitive, the detection of these lines will probably be limited to ULIRGs (unless chemical
evolutionary effects favor the emission of these oxygen lines, as discussed below).

The far-IR emission inferred from the mm/submm thermal continuum SED, along with the redshift inferred
from the CO, [CII] and [OI] lines, will allow ALMA to provide a self-consistent, unbiased view of
the evolution of the star formation rate through the cosmic epochs.
For what concerns high-z AGNs, ALMA will allow us to directly trace the coevolution of black hole
growth (through the X-ray and optical emission) and of the formation of stellar mass in the host
galaxies (through the far-IR emission). As discussed in \S
\ref{sec_past_agn_highz},
current mm/submm studies have found tentative indications of
this co-evolution for very luminous systems ($\rm SFR\sim 1000~M_{\odot}~yr^{-1}$); ALMA will extend this
investigation to much more quiescent systems ($\rm SFR\sim 10~M_{\odot}~yr^{-1}$), more representative
of the bulk of the galaxy population at high redshift. The luminosity of the
CO emission will provide the molecular gas mass for most of these galaxies and, when compared with the
SFR, will provide a direct measure of the star formation efficiency (i.e. the
star formation per unit gas mass) as a function of redshift.

In high redshift galaxies (z$>$1) the angular resolution of ALMA will allow us to resolve morphologies
on sub-kpc scales. Besides providing the extension of the star formation activity and the distribution
of the molecular gas, the ALMA maps will deliver precious information on the gas dynamics. By resolving
rotation curves it will be possible to constrain the dynamical mass of individual galaxies, which can
be compared with the expectations by hierarchical models of galaxy evolution. In the case of QSOs and type~1
AGNs it will be possible to directly compare the galaxy dynamical mass to the
black hole mass (inferred from the broad optical/UV emission lines), therefore tracing any 
putative evolution of
the $\rm M_{BH}-M_{bulge}$ relation. The observed evolution of the latter relation as a function of redshift
will be directly comparable with the predictions of models dealing with the galaxy-BH co-evolution
\citep{dimatteo05,granato04,dotti07}, and will therefore provide tight constraints on the same models.

The dynamics and morphology of the molecular gas will also provide information on the AGN fuelling mechanism.
In \S\ref{sec_past_agn_local}
we saw that local AGNs do not show obvious systematic signatures of nuclear fuelling
from the host galaxy, but also that such low luminosity AGNs do not actually require large fuelling
from the host galaxy to maintain their low accretion rates. The fuelling demand from the host galaxy
is however much larger in powerful QSOs, whose accretion rates may exceed a few $\rm M_{\odot}~yr^{-1}$.
Therefore, it is important to trace the dynamics of the molecular gas in the host of these powerful
QSOs, and investigate for instance whether in these systems gas funneled by bars or driven by galaxy
merging is common. However, such luminous QSOs are not found in the local universe, while are abundant
at high redshift. Current facilities are in general unable to resolve the molecular gas dynamics in
such distant systems, with very few exceptions. ALMA will allow us to trace in detail the gas
dynamics in a large number of QSO hosts at high redshift and to statistically investigate the occurrence
of various possible fuelling mechanisms.
Detailed simulations are already being performed to assess the detectability and capability of resolving
molecular gas morphologies in high-z AGNs. \cite{kawakatu07} showed that not only ALMA will be able to
resolve and detect the molecular gas in the host galaxy, but also the 100~pc gaseous torii expected
to surround most of the supermassive black holes at z$\sim$2.

\begin{figure}
\centerline{
\includegraphics[scale=0.6]{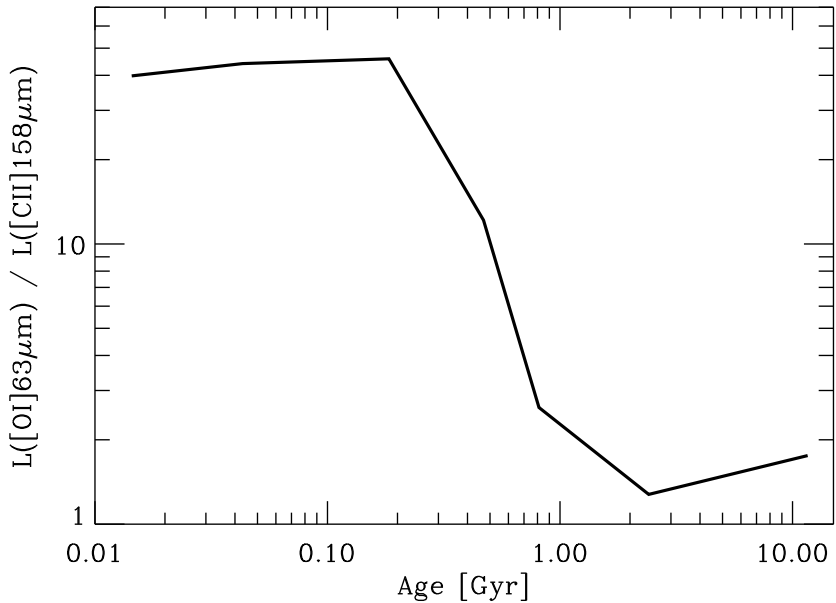}}
\caption{Expected variation of the [OI]63$\mu$m to [CII]158$\mu$m intensity ratio as a function of galaxy
age, by assuming the chemical evolutionary model in \cite{pipino04}. The model also assumes
log(U)=-2.5, $\rm n=10^3~cm^{-3}$ and that the excitation conditions do not
change over time. Courtesy of M.~Kaufman.}
\label{fig_kauf}
\end{figure}

\begin{figure}
\centerline{
\includegraphics[scale=0.67]{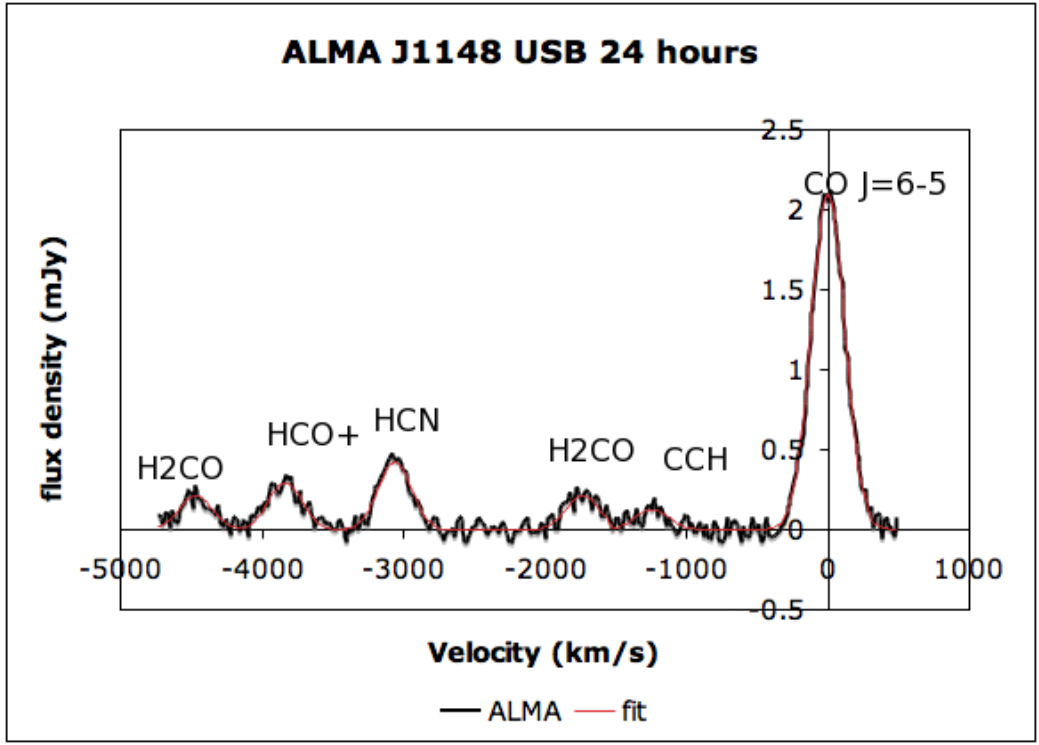}}
\caption{Simulated ALMA spectrum (24 hours of integration) of a
quasar with the same redshift (z=6.4) and luminosity of the
QSO J1148+5251, one of the most distant currently known. Courtesy of
A.~Wotten.}
\label{fig_wotten}
\end{figure}

The ALMA detection of multiple molecular and atomic lines in high-z galaxies and QSOs will also 
provide precious information on the chemical evolutionary status of the galaxy. For instance, the
ratio of [OI]63$\mu$m and [CII]158$\mu$m is sensitive to the relative abundance of oxygen and carbon C/O.
Since carbon is subject to a delayed enrichment with respect to oxygen (by several 100~Myr), the C/O
abundance ratio is a sensitive tracer of the evolutionary stage of a galaxy. Fig.~\ref{fig_kauf} shows the expected
$\rm L_{[OI]63\mu m}/L_{[CII]158\mu m}$ ratio as a function of the age of a galaxy, by exploiting the
chemical models in \cite{pipino04} (the model assumes the physical conditions derived for the QSO at z=6.4,
as discussed in \cite{maiolino05}, and that the excitation conditions do not change over time).
Therefore, by measuring both these lines, ALMA will allow us to constrain
the evolutionary stage of distant galaxies. In particular, it will be extremely interesting to investigate
the C/O abundance ratio in AGNs and galaxies at z$\sim$7, close to the re-ionization epoch, where the
age of the universe is close to the minimum enrichment timescale for carbon; at these redshifts
the C/O ratio is expected to be very low, and therefore the [OI]/[CII] ratio is
expected to be very high.

In very luminous high-z sources, such as the QSOs already detected with current facilities, ALMA will be
able to easily detect several other molecular (e.g. HCN, HCO$^+$,...) and atomic (e.g. [NII]122$\mu$m)
transitions, which will allow us to perform an accurate modelling of the physics of the
ISM and to infer accurate chemical abundances (hence the evolutionary stage of the system).
As an example, Fig.~\ref{fig_wotten} shows the simulated ALMA spectrum (24 hours of integration)
of a quasar with the same redshift (z=6.4) and luminosity of the
QSO J1148+5251, which is currently the most distant QSO with a CO detection. Beside
the CO(6--5) transition, the 8~GHz ALMA band is expected to reveal several
other molecular transitions with excellent signal-to-noise.

Finally, ALMA will allow us to investigate in detail the evolution of dust production in
the early universe. The dust mass in galaxies can only be inferred by measuring
the (rest-frame) infrared to submm SED. Currently dust masses have been measured only
for very powerful systems at high-z (QSOs and SMGs), providing an incomplete and very biased view,
not representative of the global evolution of dust through the cosmic epochs.
ALMA will measure the dust mass in large samples of high-z systems, even in relatively
quiescent ones (SFR of a few $\rm M_{\odot}~yr^{-1}$). Therefore, it will be possible to trace
the evolution of dust mass as a function of redshift and for different classes of galaxies, which
can be directly compared with models of dust evolution \citep{morgan03,dwek07,calura08}. 
In particular, it will be important to clarify whether dust produced by SNe can actually be the
main source of dust in the early universe, or whether other mechanisms of dust production are required
\citep{dwek07,elvis02}.

I am grateful to the organizers of the school for their kind invitation.
I thank M.~Walmsley, P.~Caselli, K.~Wada and L. Testi for many
useful comments on the manuscript.
I am grateful to C.~Dowell, A.~Beelen, A. Marconi, M. Kaufman and
A. Wotten for providing some of the figures in these lecture notes.
Some of the images were reproduced with kind permission
of the Space Telescope Science Institute and of the European Southern
Observatory.
Finally, I acknowledge financial support
by the National Institute for Astrophysics (INAF)
and by the Italian Space Agency (ASI).

\end{document}